\documentclass[aps,prd,preprintnumbers,notitlepage,showpacs,superscriptaddress]{revtex4-2}
\pdfoutput=1
\usepackage[usenames,dvipsnames]{color} 
\usepackage{enumitem}
\usepackage{caption} 
\usepackage{graphicx,slashed,xcolor,multirow,bbold,mathtools,sidecap,tikz,bm,ulem,enumitem,booktabs,array,amsmath,amssymb,adjustbox,setspace,caption}
\usepackage{footmisc}
\graphicspath{{Final_Plots_Resub/}}

\usepackage[colorlinks=true,linktoc=page,linkcolor=purple,citecolor=purple,urlcolor=magenta]{hyperref}
\usepackage{subcaption}
\usepackage{tikz-feynman}

\usepackage{orcidlink}
\makeatletter
\def\p@subsection{}
\makeatother

\definecolor{darkred}{rgb}{0.6,0,0}
\definecolor{denim}{rgb}{0.08, 0.38, 0.74}

\definecolor{linkcolor}{rgb}{0,0,0.5}

\def\gsim{\raise0.3ex\hbox{$\;>$\kern-0.75em\raise-1.1ex\hbox{$\sim\;$}}}
\def\lsim{\raise0.3ex\hbox{$\;<$\kern-0.75em\raise-1.1ex\hbox{$\sim\;$}}}

\def\beqn#1{\begin{equation}\label{#1}}
	\def\eeqn{\end{equation}}

\newcommand {\ignore}[1]{}

\def\321{$\mathrm{SU(3) \otimes SU(2) \otimes U(1)}$ }

\newcommand{\AddrHBNI}{
	Homi Bhabha National Institute, BARC Training School Complex, Anushakti Nagar, Mumbai 400094, India }
\newcommand{\AddrIOP}{
	Institute of Physics, Sachivalaya Marg, Bhubaneswar 751005, India}
\newcommand{\AddrIPPP}{Institute for Particle Physics Phenomenology
Department of Physics
Durham University
Durham, DH1 3LE, U.K}
\newcommand{\Addrsun}{School of Physics and Astronomy, Sun Yat-sen University, 519082 Zhuhai, China}
\newcommand{\Addrosaka}{Graduate School of Information Science and Technology, Osaka University,
Suita, Osaka 565-0871, Japan}

\definecolor{linkcolor}{rgb}{0,0,0.5}

\begin{document}

\title{\boldmath Probing right-handed neutrinos via tri-lepton signals at the HL-LHC}

\author{Manimala Mitra\orcidlink{0000-0002-8032-5125}}\email{manimala@iopb.res.in}
\affiliation{\AddrIOP}
\affiliation{\AddrHBNI}

\author{Subham Saha\orcidlink{0009-0009-1183-3271}}\email{subham.saha@iopb.res.in}
\affiliation{\AddrIOP}
\affiliation{\AddrHBNI}

\author{Michael Spannowsky\orcidlink{0000-0002-8362-0576}}\email{michael.spannowsky@durham.ac.uk}
\affiliation{\AddrIPPP}

\author{Michihisa Takeuchi\orcidlink{0000-0002-2565-7610}}\email{takeuchi@mail.sysu.edu.cn}
\affiliation{\Addrsun}
\affiliation{\Addrosaka}

\begin{abstract}
	Neutrino oscillation experiments have provided direct evidence for the existence of neutrino masses. The seesaw mechanism explains the smallness of these masses through the introduction of  heavy right-handed neutrino (RHN) states. The RHN states can aslo generate Dirac neutrino masses at tree or loop level. These heavy states can exist at the electroweak scale, approximately in the $\mathcal{O}(\mathrm{GeV})$ range, and can be investigated through current and future collider experiments. This scenario, where other new physics interactions occur at scales much higher than the RHN scale, can be described using an effective field theory (EFT) framework known as $N_R$-EFT. This study focuses on constraining the Wilson coefficients of $N_R$-EFT operators, which primarily contribute to tri-lepton production and missing energy signals at the LHC. We examine both the scenarios where the RHN mass $M_N$ is less than and greater than the $W$ boson mass $M_W$, and provide predictions for the High-Luminosity run of the LHC (HL-LHC).
\end{abstract}

\preprint{IPPP/24/33, IOP/BBSR/2024-05}

\maketitle

\section{\label{sec:1} Introduction}

Despite its remarkable predictive power, the Standard Model (SM) encounters challenges when confronted with various experimental observations. One of the most significant issues is the presence of neutrino mass and mixing, which has been experimentally verified by numerous neutrino oscillation experiments \cite{Esteban:2020cvm, deSalas:2020pgw}. This necessitates considering physics beyond the Standard Model (BSM). A subtle way to incorporate this small neutrino mass within the SM is by introducing the right-handed counterpart $N_R$ of the SM left-handed neutrino. The addition of the right chiral component can generate a small Dirac neutrino mass through the Higgs mechanism. To achieve $m_\nu \sim \mathcal{O}(0.1)$ eV, we would require the Yukawa coupling to be of $\mathcal{O}(10^{-12})$, which is philosophically unappealing. This fine-tuning can be allevaited by considering the generation of Dirac neutrino masses at the loop level \cite{Kanemura:2011jj, Yao:2017vtm, Jana_2020}.

An alternative explanation is to consider neutrinos as Majorana particles, where their tiny masses are originated from a large lepton number violating scale, as seen in the Type-I seesaw mechanism \cite{Minkowski:1977sc,Gell-Mann:1979vob,Mohapatra:1979ia,Yanagida:1979as,Yanagida:1980xy}. In the canonical Type-I seesaw scenario, the SM is extended by at least two charge-neutral SM gauge singlet Majorana fermions, $N$. These BSM particles possess a large lepton number violating Majorana mass term $M_N \overline{N^c} N$ and interact with the SM neutrinos through Yukawa interactions. The small observed neutrino mass can be realised by $m_\nu \sim y^2_\nu v^2 / M_N$, where $v$ is the SM Higgs vacuum expectation value. For $\mathcal{O}(1)$ Yukawa couplings to achieve $m_\nu \sim \mathcal{O}(0.1)$ eV, $M_N$ must be approximately $\mathcal{O}(10^{15})$ GeV, near the grand unification theory (GUT) scale.

A major challenge of this scenario is its collider testability. The large mass scale results in highly suppressed phenomenological observables. For example, right-handed neutrino (RHN) production at colliders is suppressed by the active-sterile mixing $\Theta \sim y_{\nu}v/M_N$, assuming $\mathcal{O}(1)$ Yukawa coupling. However, various low-scale seesaw mechanisms, such as inverse \cite{Mohapatra:1986aw,Mohapatra:1986bd} and linear \cite{Akhmedov:1995ip,Malinsky:2005bi}, predict pseudo-Dirac RHNs at lower mass scales that are testable at colliders, while similar testability for Dirac RHNs can be observed for the models mentioned in \cite{Kanemura:2011jj, Yao:2017vtm, Jana_2020}. These lighter RHN states can also be considered light degrees of freedom alongside SM particles within a generalized bottom-up Effective Field Theory (EFT) framework, referred to as $N_R$-EFT \cite{DELAGUILA2009399,PhysRevD.80.013010}.

Several publications address different aspects of $N_R$-EFT. References \cite{DELAGUILA2009399,PhysRevD.80.013010,PhysRevD.94.055022, PhysRevD.96.015012} and \cite{Li:2021tsq} present the non-redundant operator basis up to dimensions seven and nine of $N_R$-EFT, respectively. Collider phenomenology concerning dimension five $N_R$-EFT at future Higgs factories and the LHC is discussed in \cite{Barducci:2020icf, Caputo:2017pit, Delgado:2022fea, Duarte:2023tdw, Jones-Perez:2019plk}. Various studies \cite{delAguila:2008ir, PhysRevD.80.013010, Alcaide:2019pnf, Beltran:2022ast, Duarte:2016miz} examine subsets of higher-dimensional operators and their phenomenological implications at the LHC. Notably, if the total decay width of the light RHN is small, it can lead to interesting displaced decay signatures \cite{Drewes:2019fou, Cottin:2021lrq, Liu:2019ayx, Abada:2018sfh}. Additionally, works \cite{Barducci:2022hll, Cottin:2021lzz, Beltran:2021hpq} focus on production modes induced by different four-Fermi operators constructed at dimension six, assuming relevant decay modes for the $N$ field such as $N \rightarrow \nu \gamma$ and $N \rightarrow 3f$ (where $f$ is an SM fermion). Theoretical aspects of dimension six operators involving the Higgs doublet and their sensitivity under various Higgs-mediated processes are discussed in \cite{Butterworth:2019iff}, while \cite{Duarte:2018xst, Zapata:2022qwo} study the sensitivity of different dimension six operators at both LHC and lepton colliders.

In this work, we aim to explore and constrain few of the effective operators of $N_R$-EFT with mass dimension $d = 6$. These operators are crucial for single-producing RHNs at the LHC. We conduct a detailed collider analysis, focusing on two operators: $\mathcal{O}_{HNe}$ and $\mathcal{O}{duNe}$, and constrain the respective Wilson coefficients: $C_{\Lambda}$ and $C^\prime_{\Lambda}$. Our analysis primarily focuses on the leptonic decay $N \rightarrow e \ell \nu$ of the RHN, produced via Drell-Yan (DY) interaction at the LHC: $p p \rightarrow e^\pm N$, leading to a final state containing tri-lepton.

This paper is organised as follows: Section \ref{sec:2} briefly discusses our theory setup and the branching ratios of the RHN. In Section \ref{pptolNcrossx}, we discuss the production cross-section of tri-lepton+MET. Section \ref{sec:4} covers various constraints on the respective Wilson coefficients. Section \ref{sec:5} provides a detailed collider analysis for the HL-LHC and presents our results. Finally, Section \ref{conclusion} summarises our findings.

\section{\label{sec:2} Theoretical set-up}

The main thrust of this paper is to portray the importance of a few of the $d=6$ operators in the $N_R$-EFT framework, which enhances RHN production at the LHC and at its high-luminosity run. We, therefore, do not adhere to any specific neutrino mass model, rather consider a SM gauge singlet fermionic state  $N$ with mass $M_N > \mathcal{O}(10)$ GeV that mixes with the SM neutrino state $\nu_e$ with active-sterile mixing $\chi$. For simplicity, in writing the charged current interaction, we only consider an electron. The charged  current interaction term that originates from renormalisable Lagrangian has the following form \cite{Banerjee:2015gca}
\begin{equation}
\mathcal{L}_{lNW}=\frac{g}{\sqrt{2}}W^-_{\mu}\bar{e}\gamma^{\mu}P_L \chi N+H.C
\label{eq:renormNRint}
\end{equation}
 Generalisation of the above equation to include different leptons is straightforward. In addition to this, $N$ will also have interaction, such as $N-\nu_e-Z$ and $N-\nu_e-H$. We refer the reader to \cite{Banerjee:2015gca} for the details. Embedding the RHN state in an EFT framework enables a new set of operators containing $N$  and the SM fields. A summary of the $d=5$ and $d=6$ operators involving RHN state can be found in \cite{Mitra:2022nri}. Among them, our specific interests lie in the operators which can enhance single RHN production at the LHC and HL-LHC. We hence focus on $\mathcal{O}_{HNe}$ and $\mathcal{O}_{duNe}$ operators. 
$\mathcal{O}_{HNe}$ gives  rise to the production of $N$ from the decay of $W$ gauge boson and has the following form,
	\begin{equation}
		\mathcal{O}_{HNe} = \frac{c}{\Lambda^2} (\overline{N}_R \gamma^{\mu}e_R)({\tilde{H}}^\dagger i D_\mu H) + H.C,
		\label{eq:hne}
	\end{equation}
	 where $N_R$ is the right-chiral component of  $N$. The  non-renormalizable four Fermi operator $\mathcal{O}_{duNe}$ involving the RHN state $N_R$ has the following form, 
	\begin{equation}
		\mathcal{O}_{duNe}=\frac{c^{\prime}}{\Lambda^2}\left(\bar{d}_R\gamma^{\mu}u_R\right)\left(\bar{N}_R\gamma_{\mu}e_R\right) + H.C
		\label{eq:duNe}
	\end{equation}
Other $d-6$ four-Fermi operators such as $\mathcal{O}_{LNQd,LdQN,QuNL}$ \cite{Mitra:2022nri} will have a similar effect in the production of $N$. We here follow a minimalistic approach and consider only $\mathcal{O}_{duNe}$.
	The  dimension-6 operator given in Eq.~\ref{eq:hne}  gives rise to the following interaction term of $N$ with the SM gauge boson $W$,
	\begin{equation}
		\mathcal{L}=\frac{g}{\sqrt{2}}W^-_{\mu}\bar{e}\gamma^{\mu}  {\Theta}  N+H.C
		\label{eq:nreft1}
	\end{equation}
	where the coefficient coupling $\Theta$ can be expressed as follows, 
	\begin{equation}
		\Theta = \chi P_{L} + \frac{ c v^{2}}{2\Lambda^{2}}P_{R}.
		 \label{eq:vfc}
	\end{equation}

\begin{figure}[ht]
	\centering
	\begin{minipage}{0.50\textwidth}
		\centering
		\begin{tikzpicture}
			\begin{feynman}
				\vertex (a);
				\vertex[above left=of a](b);
				\vertex[below left=of a](c);
				\vertex[blob, draw=blue, fill=blue, minimum size=4mm, right=of a] (f) {};
				\vertex[above right=of f](d);
				\vertex[blob, draw=blue, fill=blue, minimum size=4mm, below right=of f] (e) {};
				\vertex[above right=of e](h){$e^{\pm}$};
				\vertex[right=of e](k);
				\vertex[above right=of k](x);
				\vertex[below right=of k](y);
				\diagram*  {
					(b) -- [edge label=$q$, fermion] (a),
					(c) -- [edge label=$\bar{q^{\prime}}$, fermion] (a),
					(a) -- [edge label=$W^{\pm}$, boson] (f),
					(f) -- [edge label=$e^{\pm}$, fermion] (d),
					(f) -- [edge label=$N$, fermion] (e),
					(e) -- [fermion] (h),
					(e) -- [boson] (k),
					(k) -- [ fermion] (x),
					(k) -- [edge label=$\nu$, fermion] (y),
				};
			\node at (5.6, -0.9) {${\ell}^{\mp}(e^{\mp}, \mu^{\mp})$};
			\node at (3.8, -0.8) {$W^{\mp}$};
			\end{feynman}
		\end{tikzpicture}
	\end{minipage}%
	\hfill
	\begin{minipage}{0.50\textwidth}
		\centering
		\begin{tikzpicture}
			\begin{feynman}
				\vertex[blob, draw=red, fill=red, minimum size=4mm, right=of a] (a) {};
				\vertex[above left=of a](b);
				\vertex[below left=of a](c);
				
				\vertex[above right=of a](d);
				\vertex[blob, draw=blue, fill=blue, minimum size=4mm, below right=of a] (e) {};
				\vertex[above right=of e](h){$e^{\pm}$};
				\vertex[right=of e](k);
				\vertex[above right=of k](x);
				\vertex[below right=of k](y);
				\diagram*  {
					(b) -- [edge label=$q$, fermion] (a),
					(c) -- [edge label=$\bar{q^{\prime}}$, fermion] (a),
					(f) -- [edge label=$e^{\pm}$, fermion] (d),
					(f) -- [edge label=$N$, fermion] (e),
					(e) -- [fermion] (h),
					(e) -- [boson] (k),
					(k) -- [ fermion] (x),
					(k) -- [edge label=$\nu$, fermion] (y),
				};
			\node at (5.6, -0.9) {${\ell}^{\mp}(e^{\mp}, \mu^{\mp})$};
			\node at (3.8, -0.8) {$W^{\mp}$};
			\end{feynman}
		\end{tikzpicture}
	\end{minipage}
	\caption{Feynman diagrams for signal process of $p p \to e N(N\rightarrow e \ell \nu)$. As stated in the text,  the red and blue blobs represent the four-Fermi operator $\mathcal{O}_{duNe}$ and a combination of $\chi$ and $\mathcal{O}_{HNe}$  respectively.}
	\label{Fig:NR_production}
\end{figure}
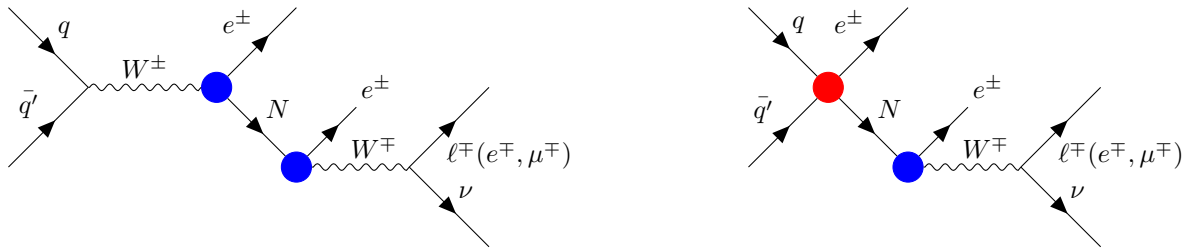

From Eq.~\ref{eq:nreft1}, this is evident that $O_{HNe}$ can directly influence $N \rightarrow e W$ and $N \rightarrow e W^{*} \rightarrow e l \nu$ decays, for $M_{N} > M_{W}$ and $M_{N} < M_{W}$ respectively. The production of $N$ in association with $e^{\pm}$, followed by its subsequent decay via the process $N \to e^{\pm} l^{\mp} \nu$, is illustrated in FIG.~\ref{Fig:NR_production}. The blue blob represents the effective $WeN$ vertex given in Eq.~\ref{eq:nreft1}, while the red blob represents the four fermi $\mathcal{O}_{duNe}$ vertex. Besides affecting the $p p \to e N$ production, both operators have an impact in modifying the branching ratio of $N$, which we discuss in detail. In this work, we pursue an in-depth study of tri-lepton associated with missing energy, one of the most prominent channels for the RHN search. The operator $\mathcal{O}_{duNe}$ will not directly appear in the $N$ decay diagram for the $N \to e\ell \nu$ final state. However, the branching ratio of $N \to e\ell \nu$ heavily relies on both the operators, as due to the presence of $\mathcal{O}_{duNe}$, decay channel $N \to e q \bar{q^{\prime}}$ can have large branching ratios. For the subsequent discussion, we consider  $\chi=10^{-5}$, which is closer to the value in agreement with the light neutrino mass constraint for Type-I seesaw. Before proceeding further, we redefine the respective Wilson coefficients of 
$\mathcal{O}_{HNe}$ and $\mathcal{O}_{duNe}$ as follows, 
\begin{equation}
C_{\Lambda}=\frac{c}{\Lambda^2}, ~~~~
C^{\prime}_{\Lambda}=\frac{c^{\prime}}{\Lambda^2}
\label{eq:wilredine}
\end{equation}
Accordingly, the coupling strengths of the red and blue blobs become propotional to $\left( \chi P_{L} + \frac{C_{\Lambda}  v^{2}}{2}P_{R} \right)$ and $C'_{\Lambda}$, respectively. Note that, as indicated in Eq.~(\ref{eq:wilredine}), both $C_{\Lambda}$ and $C'_{\Lambda}$ have units of $\text{GeV}^{-2}$, which will not be reiterated in the subsequent discussions.


\subsection{Branching Ratios of Different Decay Modes of \texorpdfstring{$N$}{} \label{sec:branchingnr}}

\begin{figure}[h!]
	\centering
    \includegraphics[scale=0.55]{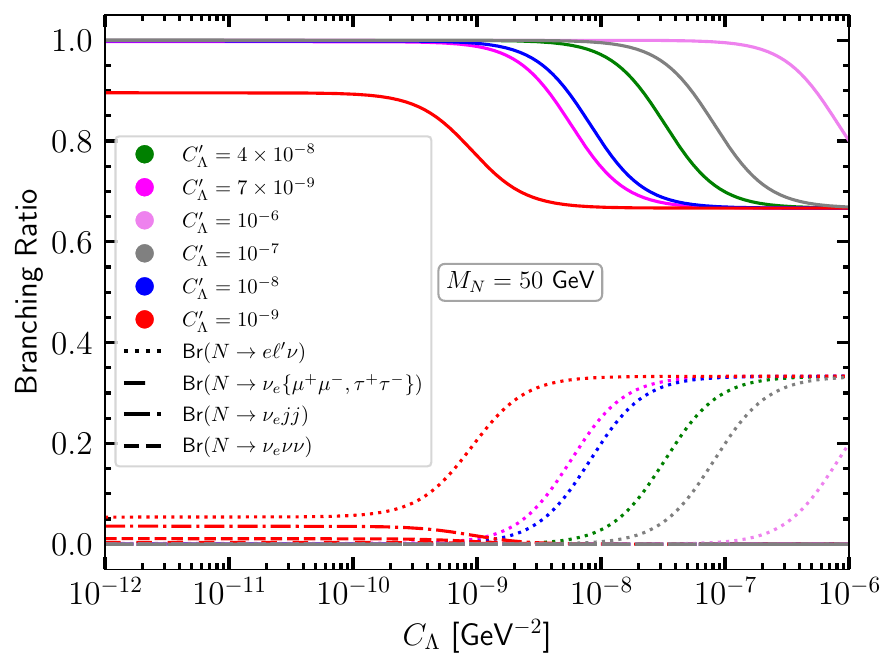}
	\includegraphics[scale=0.55]{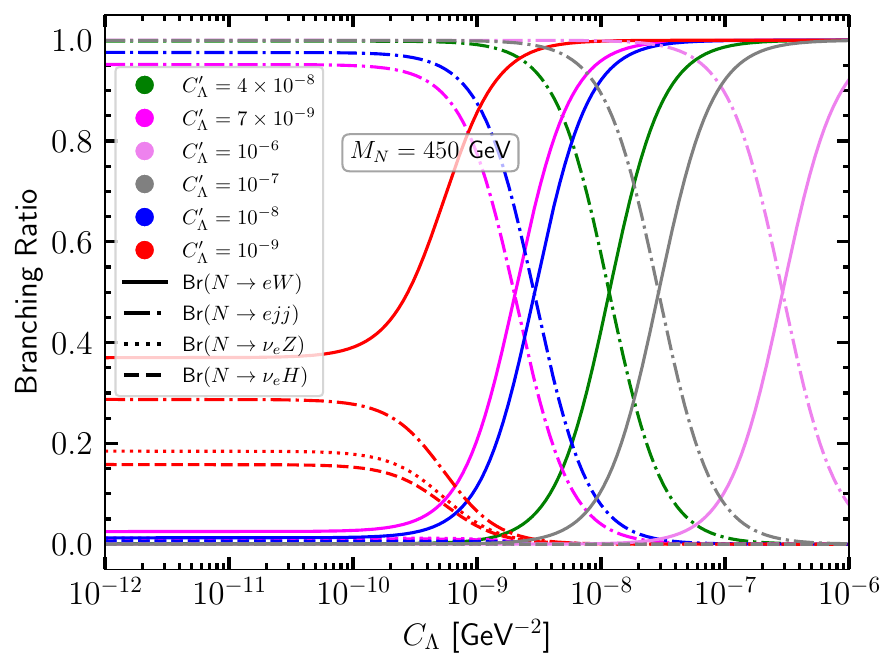}
	\captionsetup{justification=centering}
	\caption{Branching ratios of $N$ for $M_N=50$ GeV (left panel) and for $M_N=450$ GeV (right panel).}
	\label{Fig:BR_NR}
\end{figure}

The RHN can decay to a number of final states, $N \to \{e l^{\prime} \nu\}$ (mediated by $W$), $\{e j j\}$ (mediated by $W, C^{\prime}_{\Lambda}$), $\{\nu_e j j,  \nu_e \nu \nu\}$ (mediated by $Z$), $\{\nu_e {l^{\prime}}^+{l^{\prime}}^-,  \nu_e c \bar{c}, \nu_e b\bar{b}\}$ (mediated by $Z, H$) for $M_N < M_W$ and to states $N \to \{e W, ej j\}$ for $M_N> M_W$, $\{ e W, \nu_e Z, ej j\}$ for $M_N> M_Z$    and $\{e W, \nu_e Z,  \nu_e H, e j j\}$ 	for $M_N> M_H$. The $j$ in the above represents light quarks and ${\ell}^{\prime}$ represents $e , \mu, \text{and} $ $\tau$. When analyzing the model signature in Section~5, we consider the semileptonic decays of $\tau$. The decay mode  $ejj$ has a substantial contribution from $C^{\prime}_{\Lambda}$, and for a larger coupling, $ejj$ mode can even be larger than the two-body modes. Note that, for $M_W<M_N < M_Z$, there will be off-shell contribution such as $N \to \nu_e Z^* \to \nu ab$, where $a,b$ are different SM fermions. However, this off-shell contribution is suppressed compared to $eW, ej j$. A similar conclusion also holds for three body decays via  Higgs mediation when $M_Z < M_N < M_H$. We derive the analytic expressions of partial decay widths of $N$ for all possible final states and provide the expressions in the Appendix.  In FIG.~\ref{Fig:BR_NR}, we show the variation of branching ratios with  $C_{\Lambda}$ for different values of  $C^{\prime}_{\Lambda}$ for $M_N=50$ GeV (see left panel) and for $M_N=450$ GeV (see right panel) for our chosen value of activate-sterile mixing $\chi= 10^{-5}$. From these two figures, it is evident that at higher $M_N$ values, $N$ decaying to two-body states primarily occur, along with the three-body decay $ejj$ via $C_{\Lambda}^{\prime}$, whereas at lower masses, three-body decays $N \to e l^{\prime} \nu, \nu_e \{ \mu^{+} \mu^{-} , \tau^{+} \tau^{-} \}   , e j j, \nu_e j j, \nu_e \nu \nu $ prevail.  A few comments are in order:

\begin{figure}[t!]
	\centering

	\includegraphics[scale=0.6]{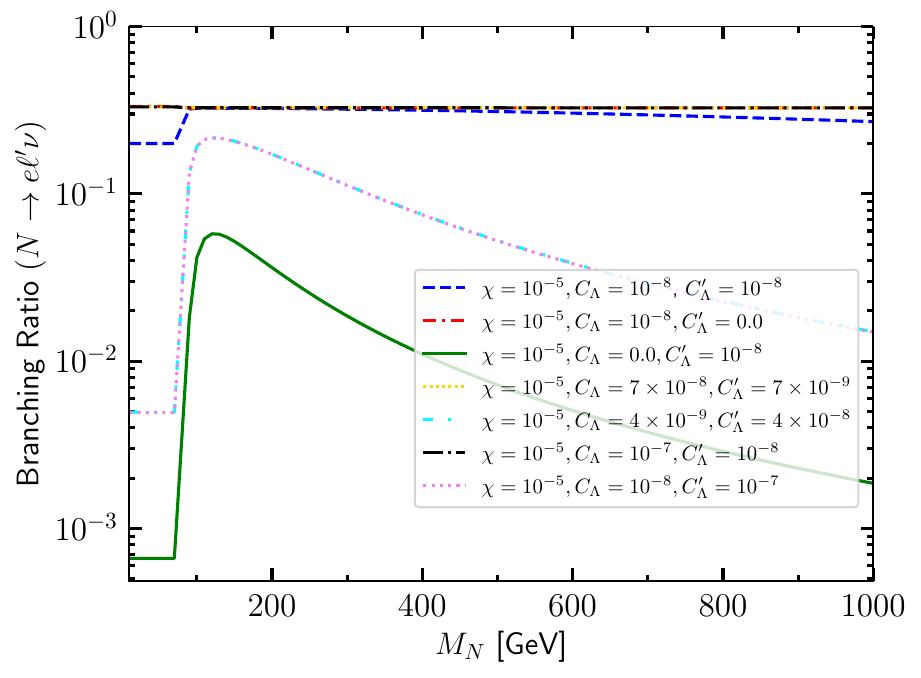}
	\caption{Branching Ratio of $N \rightarrow e l^{\prime} \nu$  with varying $M_N$. The different coloured lines represent the branching ratio for different values of the Wilson coefficients.}
	\label{Fig:Cross_Section_BR_elv}
\end{figure}

\begin{itemize}
	\item For $M_N=50$ GeV, the branching ratios for all the decay channels remain approximately constant for  $10^{-12} < C_{\Lambda}<  10^{-10}$. Since in this region,  $C_{\Lambda}$ is significantly smaller compared to $\chi$, in the vertex factor $\left( \chi P_{L} + A P_{R} \right) \approx \chi P_{L}$, where $A = \frac{  v^{2}}{2} C_{\Lambda}$, hence the partial decay widths of different channels, and branching ratios are nominally dependent on $C_{\Lambda}$. Therefore, the branching ratios of different modes are dictated by simply two parameters $\chi$ and $C^{\prime}_{\Lambda}$. We can see in this range that the branching ratio of $e j j$ is significantly higher for our choice of $C^{\prime}_{\Lambda}$. Lowering down $C^{\prime}_{\Lambda}$ from $10^{-6}$ to $10^{-9}$ reduces the branching ratio of $e j j$  from 100$\%$ to approximately 89$\%$  and enhances the branching ratio of other modes which are solely governed by $\chi$. In this range, the maximum value of the branching ratio for the $e l^{\prime} \nu$ and $\nu_e j j$ modes can slightly exceed $5\% $. 
	
	
	\item Exceeding $C_{\Lambda} = 10^{-10}$, the contribution of $C_{\Lambda}$ becomes more prominent than $\chi$ in the vertex factor, so that  $\left( \chi P_{L} + A P_{R} \right) \approx A P_{R}$.  In this region, a competition arises between $C_{\Lambda}$  and $C^{\prime}_{\Lambda}$. As $C_{\Lambda}$ continues to increase, the partial decay width of $e l^{\prime} \nu$ whose decay width is governed by $A$  increases, resulting in an increment in the respective branching ratio. For $ejj$, there is a contribution both from $\left( \chi P_{L} + A P_{R} \right)$ and $C^{\prime}_{\Lambda}$, and there is interference between different contributions, as can be seen from the Appendix. For this channel, there is initially a suppression in the branching ratio and at a large value of $C_{\Lambda}$, the partial decay width of this channel is completely governed by $C_{\Lambda}$, resulting in a saturated value of branching ratio. We find that the saturating  value of branching ratios for $ejj$ is  $67\%$ and 33$\%$ for $el\nu$ for $C^{\prime}_{\Lambda}=10^{-7},4 \times 10^{-8}, 10^{-8}, 7 \times 10^{-9} , 10^{-9}$.  For other modes, the branching ratios are negligible. 
	
	\item  For a mass of 450 GeV, the two body decay modes $N \to e W$ become the most dominant for a large value of $C_{\Lambda}$. As an example, for $C^{\prime}_{\Lambda}=7 \times 10^{-9}$, when $C_{\Lambda} > 2 \times 10^{-8}$, the branching ratio of $N \to e W$ is almost 100$\%$. Contrary to this, the dominance of the three body decay mode $N \rightarrow e jj$ occurs when  $\left( \chi P_{L} + A P_{R} \right)$ contribution is smaller than $C_{\Lambda}^{'}$ one. The other decay modes, such as $N \to \nu_e Z, \nu_e H$, are heavily suppressed due to the choice of a small $\chi$.

\end{itemize}

In FIG.~\ref{Fig:Cross_Section_BR_elv}, we show the variation of $N \rightarrow e l^{\prime} \nu $ branching ratio with respect to $M_N$.  For the blue dashed curve, both $C_{\Lambda}$ and $C_{\Lambda}^{\prime}$ are set to $10^{-8}$. For the red dot-dashed and green solid curves, we set $C_{\Lambda}$ to $10^{-8}$ and $C_{\Lambda}^{\prime}$ to $0$, and vice versa. 
For the black curve, we set $C_{\Lambda}$ to $10^{-7}$ and $C_{\Lambda}^{\prime}$ to $10^{-8}$, and for pink one we reverse these values for comparison. Additionally, for the yellow and cyan curves, we utilise benchmark points where $C_{\Lambda}= 7 \times 10^{-8}$,   $C_{\Lambda}^{\prime}=7 \times 10^{-9}$, and $C_{\Lambda} =4 \times 10^{-9}, C_{\Lambda}^{\prime}= 4 \times 10^{-8}$, respectively. Among these benchmark points, when $C_{\Lambda} > C^{\prime}_{\Lambda}$, the branching ratio ($N \to e l^{\prime} \nu$ )  remains constant, $\approx 33\%$, across mass values due to the negligible effect of $C_{\Lambda}^{\prime}$, thus minimising the suppression from the four-Fermi vertex as can be seen from red, yellow and black curves. For $C_{\Lambda} < C^{\prime}_{\Lambda}$, when \( M_{N} < M_{W} \)  the partial decay-width of $N \rightarrow e l^{\prime} \nu $ is suppressed due to an off-shell \( W \). Conversely, when \( M_{N} \geq M_{W} \), $N$ decays via two body decay mode to $N \to e W$, resulting in an enhancement in the partial decay widths and branching ratio of $N \to e W \to e l^{\prime} \nu$. As a result, the branching ratio for the process $N  \to e l^{\prime} \nu$ experiences a jump when \( M_{N} \) crosses \( M_{W} \), as is visible in green, cyan and pink curves. As the mass increases further, $e l^{\prime} \nu$ channel's partial decay width increases as $M^3_N$. For the $N \rightarrow e j j$ channel, on-shell $W$ contribution which is dependent on  $C_{\Lambda}$  is  smaller due to $C_{\Lambda} < C^{\prime}_{\Lambda}$. The direct three body decay $N \to e j j$, which depends on $C^{\prime}_{\Lambda}$ is more dominant. Hence,  the  partial decay width of this channel increases as $M^5_N$ resulting in a rapid increase in the $e j j$ partial width with $M_N$. Therefore, the branching ratio of the $e l^{\prime} \nu$ channel falls off with increasing $M_{N}$ after surpassing \( M_{W} \).  For \(C_{\Lambda} = C'_{\Lambda}\), since the effects of both the Wilson coefficients are nearly equal, we observe an average effect, resulting in a smaller jump around \(M_N = M_W\) as depicted by the blue curve.

\section{ \texorpdfstring{$p p \to e N \to e e l^{\prime} \nu$}{} cross-section \label{pptolNcrossx}}

	The presence of $\mathcal{O}_{HNe}, \mathcal{O}_{duNe}$ operators will enhance the production cross-section of $p p \to e N$ channel as compared to the scenario, when only $\chi$ is present. To compute the cross-section, we implement the model in Feynrules \cite{Alloul:2013bka} and generate events via MadGraph5aMC@NLO -v3.5.3 \cite{Alwall:2014hca}.  The LHE file is then passed through PYTHIA v8.310 \cite{Sjostrand:2014zea} for showering and hadronisation, and we implement our analysis on this event sample. In FIG.~\ref{Fig:cross-section_pp_ln}, we show the production cross-section of $pp\rightarrow e N$ for 14 TeV c.m. energy, taking into account the decay of $N \to e l^{\prime} \nu$ final state. To evaluate the cross-section, we generate  $p p \to eN$ in MadGraph and subsequently fold the cross-section with the branching ratios of $N \to e l^{\prime} \nu$ from our analytic calculation. The chosen values of $C_{\Lambda}$ and $C^{\prime}_{\Lambda}$ are kept same as shown in FIG.~\ref{Fig:Cross_Section_BR_elv}. A few comments are in order:  
\begin{figure}[htb!]
 	\centering
 	\includegraphics[scale=0.6]{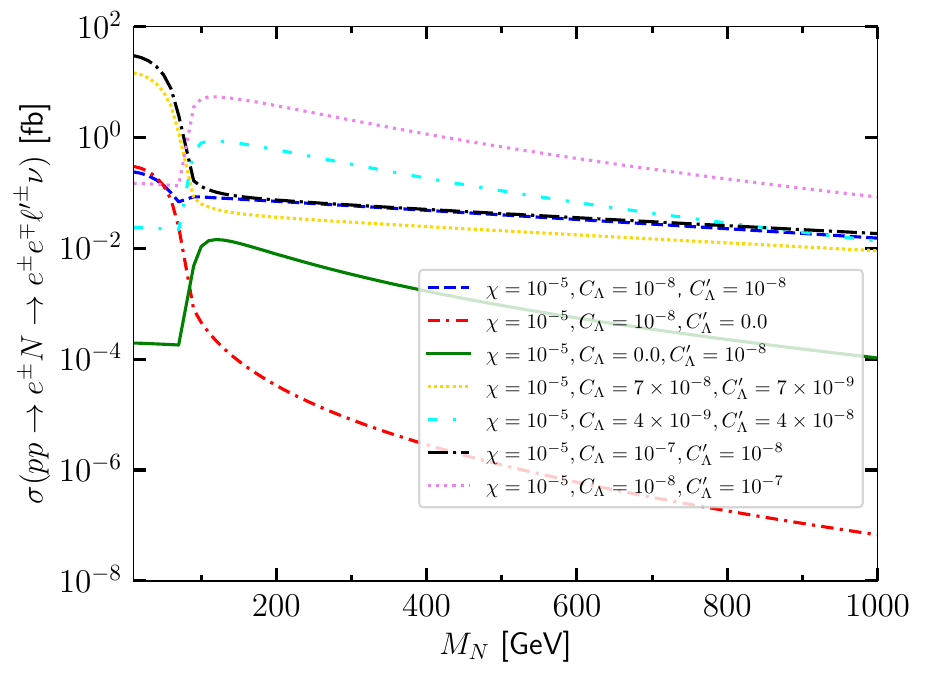}

 	\caption{Production cross-section of $pp \rightarrow e N \to e e l^{\prime} \nu $ for different $M_N$. The different coloured lines represent the total cross-section for different values of the Wilson coefficients.}
 	\label{Fig:cross-section_pp_ln}
 \end{figure}

\begin{itemize}
	\item The plotted cross section of $p p \rightarrow ee {\ell}^{\prime} \nu$ in FIG.~\ref{Fig:cross-section_pp_ln} results from the multiplication of the cross-section for the process \( p  p >e  N  \) and the branching ratio of the \( N \rightarrow e{\ell}^{\prime} \nu \) channel. 
 Naturally, the variation in the total $p p \rightarrow e e {\ell}^{\prime} \nu $ cross-section is influenced by  $p p \rightarrow e N$    cross-section and the branching ratio of $N   \rightarrow e {\ell}^{\prime} \nu$. Here $N, e, {\ell}^{\prime}$ and $\nu$  represent $N / \bar{N}$, $e^{+/-}$,  ${{\ell}^{\prime}}^{+/-}$ and  $\nu / \bar{\nu}$ respectively. 
	The $p p \rightarrow e N$ production occurs via two channels as has been shown in FIG.~\ref{Fig:NR_production}, i.e. by both four-Fermi (red blob) and $W^\pm$ mediation (blue blob). 
   
   \item 
   We note that in scenarios where $C_{\Lambda} > C'_{\Lambda}$, for  $M_N$ much smaller than $M_W$, the primary production mode is $pp \rightarrow W \rightarrow eN$. As $M_N$ approaches a threshold $M_N^{{\text{th}}_{1}}$ closer to $M_W$, the contribution from the four-Fermi operator to the $pp \rightarrow eN$ cross-section becomes relatively more significant among the two contributions, as $C_{\Lambda}$ contribution receives kinematic suppression. Beyond $M_W$, as $M_N$ increases, the four-Fermi operator's contribution continues to become more relevant in the $p p \to e N$ cross-section as $pp \rightarrow W^{*} \rightarrow eN$ process is heavily suppressed. Once $M_N$ surpasses another threshold $M_N^{{\text{th}}_{2}}$($> M_W$), the four-Fermi operator dominates the $pp \rightarrow eN$ cross-section. Conversely, when $C_{\Lambda} < C'_{\Lambda}$, the dominance of the four-Fermi operator extends across all mass regions.

	\item
	
		When \(C_{\Lambda}  > C'_{\Lambda} \) and \( M_{N} > M_{W} \), the intermediate \( W \) boson is off-shell in $pp \rightarrow eN$ and therefore the process receives propagator suppression. For \( M_{N} < M_{W} \), the produced $W$ is instead  on-shell and  decays to an $eN$ state. Hence, increasing $M_N$ from \( M_{N} < M_{W} \) to \( M_{N} > M_{W} \), the $p p \rightarrow e N$ cross section decreases as can be seen from red, yellow and black lines in FIG.~\ref{fig:cmslimit}.  With further increasing $M_N$, the cross-section becomes even more phase-space suppressed. This suppression becomes more prominent for the scenario \( C_{\Lambda}^{\prime} = 0 \) as the production is only by $W$ mediated diagram. As the $N \to e {\ell}^{\prime} \nu$ branching ratio remains constant, $\approx 33\%$ across mass values the $p p \rightarrow e e {\ell}^{\prime} \nu$ cross-section as shown in FIG.~\ref{Fig:cross-section_pp_ln} mirrors the $p p \rightarrow eN$ cross-section's behaviour.
	
	\item
	 
		 For cases where \(C_{\Lambda} < C_{\Lambda}^{\prime}\), the suppression in cross-section for the process  $ p p \to e N$ for large mass $M_N \gg M_W $ is nominal,  as is evident from the green, cyan and pink curves in FIG.~\ref{fig:cmslimit}.  
The $N \rightarrow e {\ell}^{\prime} \nu$ branching ratio experiences a jump when \( M_{N} \) crosses \( M_{W} \), as is clearly visible by the respective curves in FIG.~\ref{Fig:Cross_Section_BR_elv}. This jump translates into an enhanced $p p \rightarrow e e {\ell}^{\prime} \nu $  cross-section in FIG.~\ref{Fig:cross-section_pp_ln} around $M_N \approx M_W$ depicted by the same coloured curves mentioned previously. 

After surpassing $M_W$, the branching ratio of the $N \rightarrow e {\ell}^{\prime} \nu$ channel in this scenario falls off. For higher $ M_{N}$, as $p p  \rightarrow e   N$ cross-section and $N \rightarrow e {\ell}^{\prime} \nu$ branching ratio both decrease, the signal cross section $p p \rightarrow e e {\ell}^{\prime} \nu$ also decreases.  From FIG.~\ref{Fig:cross-section_pp_ln}, this is also  evident that for a fixed $C^{\prime}_{\Lambda}$, an increase in  \( C_{\Lambda} \)  increases the \(pp \rightarrow e e {\ell}^{\prime} \nu\) cross-section. This occurs as a higher \( C_{\Lambda} \) contributes to an increase in both \(p p \rightarrow eN\) production and the \(N \rightarrow e {\ell}^{\prime} \nu\) branching ratio.

	\item
	In cases where $ C_{\Lambda}^{\prime} = C_{\Lambda}$, the $ N \rightarrow e {\ell}^{\prime} \nu$ branching ratio depicts similar enhancement around $M_N \sim M_W$, as is visible from the blue dashed line of   FIG.~\ref{Fig:Cross_Section_BR_elv}. In this case,  the $p p \rightarrow eN$ cross-section experiences minimal kinematic suppression as \( M_N \) increases (see the blue dashed line in  FIG.~\ref{fig:cmslimit}). Due to the presence of a large $C_{\Lambda}$, the fall in the $N \rightarrow e {\ell}^{\prime} \nu$ branching is somewhat nominal as compared to $C_{\Lambda}  <  C_{\Lambda}^{\prime}$. Finally, the $ p p \rightarrow ee \ell^{\prime} \nu $ cross-section falls nominally with increasing $M_{N}$, as visible in the blue dashed line of FIG.~\ref{Fig:cross-section_pp_ln}.
\end{itemize}

\section{Existing Constraints on $N$ \label{sec:4}}

 Several searches at the LHC have been performed to investigate the presence of a right-handed neutrino state \cite{CMS:2018iaf,CMS:2018jxx, Cms:2024, ATLAS:2019kpx, CMS:2022chw, CMS:2022fut}. The CMS and ATLAS searches analysed the prompt decays of $N$ leading to di-lepton+dijet \cite{CMS:2022chw}, tri-lepton+MET signatures \cite{CMS:2018iaf,CMS:2018jxx, Cms:2024, ATLAS:2019kpx}, while searches such as 
\cite{CMS:2022fut} investigated possible displaced decay signatures. Since the final state in our case is similar to the signatures analyzed in the aforementioned searches \cite{CMS:2018iaf, CMS:2018jxx, Cms:2024, ATLAS:2019kpx}, they have the potential to constrain our model parameters. Below, we identify a few different searches of heavy neutral lepton at the LHC and analyse the impact of these respective searches on our parameters.

	\begin{figure}[htb!]
		\centering
		\includegraphics[scale=0.5]{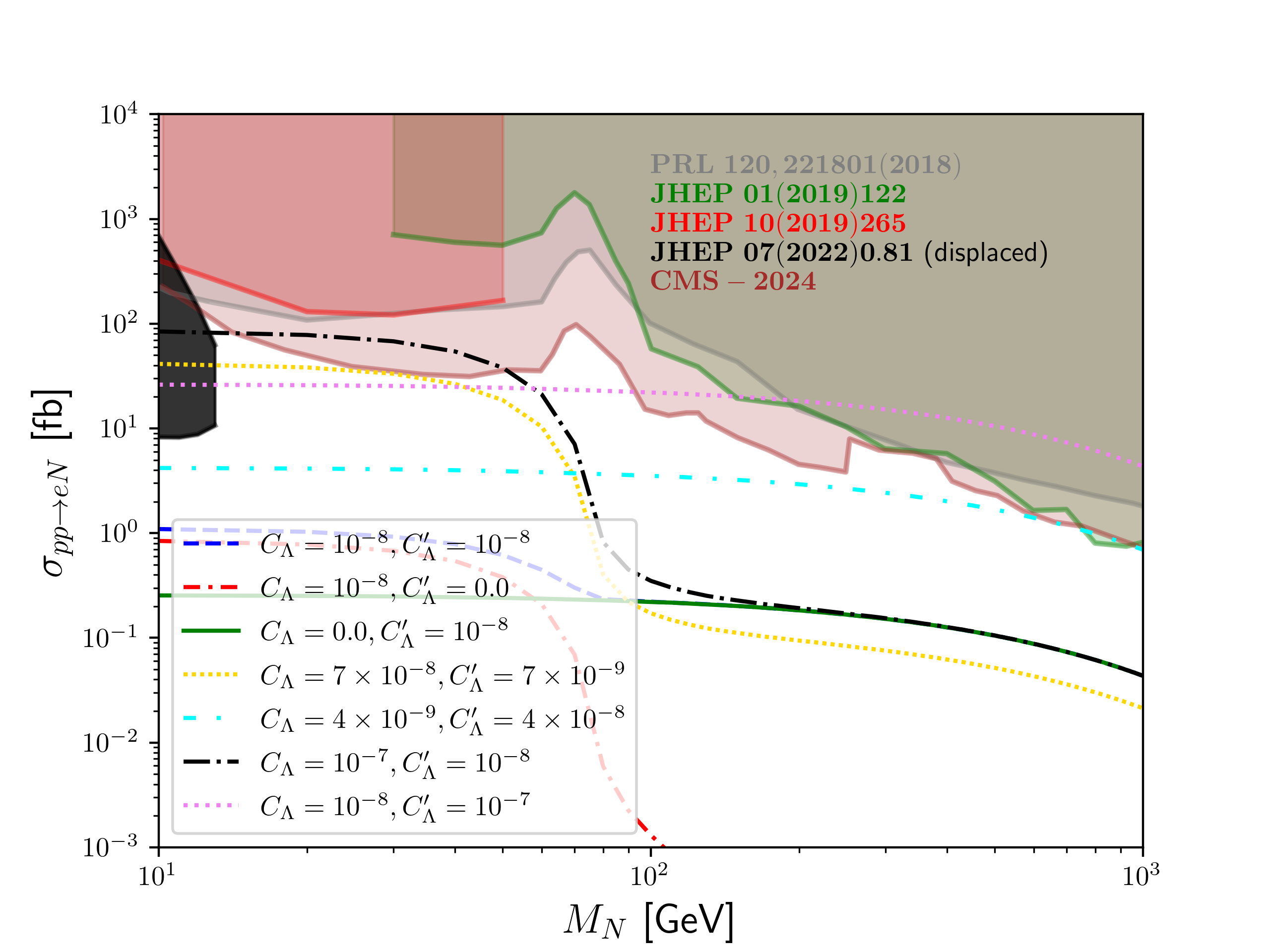}
		\includegraphics[scale=0.5]{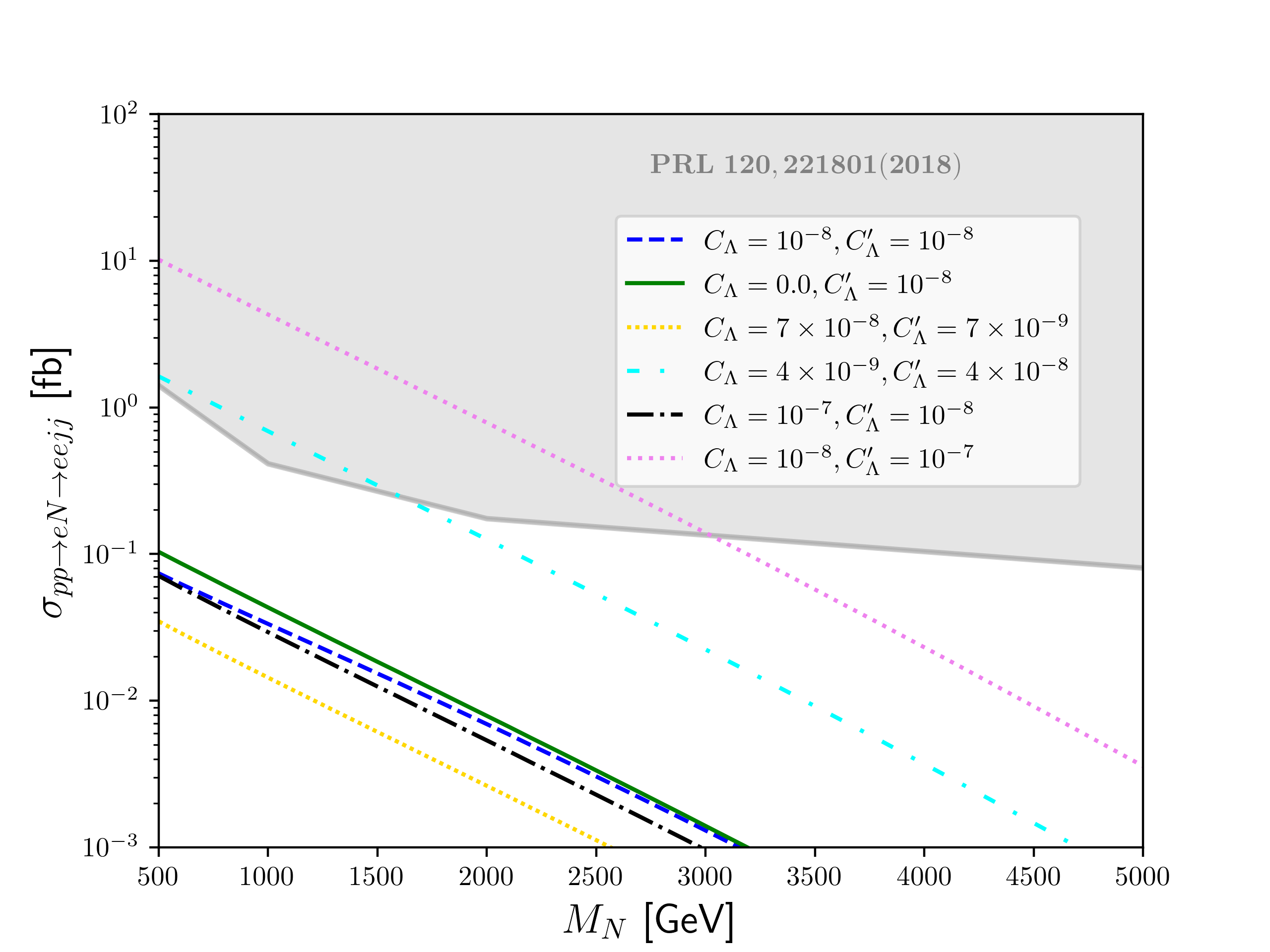}
		\caption{Left: The figure represents the total cross-section for $pp\rightarrow e N$. The grey, green and red shaded regions are the upper limit on the cross-section for the mentioned signal, reinterpreted from the result of CMS and ATLAS analysis \cite{CMS:2018iaf,CMS:2018jxx, Cms:2024, ATLAS:2019kpx}. The black-shaded region is excluded because of displaced search \cite{CMS:2022fut}. Right: The figure represents the total cross-section of $pp\rightarrow e N(N\rightarrow e j j )$ for different values of the couplings. The shaded region is the upper limit on the cross-section obtained from CMS search \cite{CMS:2022chw}.  We considered the active-sterile mixing $\chi = 10^{-5}$ in both figures.}
	\label{fig:cmslimit}
	\end{figure}
	
	\begin{itemize}
		\item In the left panel of FIG.~\ref{fig:cmslimit}, the variation of the $p p \rightarrow e N$ cross section is shown for various combinations of $C_{\Lambda}$ and $C_{\Lambda}^{\prime}$ with different $M_{N}$ values. {We kept the chosen values of the respective Wilson coefficients the same as those mentioned in FIG.~\ref{Fig:cross-section_pp_ln} and FIG.~\ref{Fig:Cross_Section_BR_elv} for ease of comparison.} The grey, green, brown and red shaded regions represent the excluded regions, obtained from the CMS and ATLAS prompt searches \cite{CMS:2018iaf,CMS:2018jxx,Cms:2024,ATLAS:2019kpx}. The black shaded region is excluded from displaced search \cite{CMS:2022fut} of RHN. We obtained these upper limits on the cross sections by considering the upper limits on the active sterile mixings presented in the experimental searches and following the assumption $C_{\Lambda} = C_{\Lambda}^{\prime} = 0$. Note that $C_{\Lambda}=0$ is not a necessary condition, as in the presence of a non-zero $C_{\Lambda}$, the overall $p p \to e N$ cross-section is rescaled as $(\chi^2 + A^{2})\sigma^2_0$ along with the interference, where $\sigma_0$ is the cross-section with $C_{\Lambda}=0$ and $\chi$ set to unity. For a non-zero $C^{\prime}_{\Lambda}$, the additional contribution scales as ${C^{\prime}_{\Lambda}}^2 \sigma_1$, along with interference terms. Here, $\sigma_1$ represents the cross-section for $p p \to eN$ governed solely by the contact four-Fermi interaction, with $C^{\prime}_{\Lambda}$ set to unity. A comparison between different searches, as depicted in the figure, shows that for light RHN, the exclusion from the displaced search is more stringent, as can be seen from the black-shaded region. Other than this, the searches performed by CMS \cite{Cms:2024,CMS:2018jxx} put stringent constraints throughout the parameter space up to $M_N = 1$ TeV. However, these searches are only one-to-one applicable if $C^{\prime}_{\Lambda}=0$, in which case only the left panel diagram shown in FIG.~\ref{Fig:NR_production} is present. In the presence of a non-zero and large $C^{\prime}_{\Lambda}$, the cut-efficiency following \cite{Cms:2024,CMS:2018jxx} is significantly reduced, as will be discussed in the following section. As a result, the effectiveness of constraints in scenarios where $C^\prime_{\Lambda}$ is higher is still uncertain, and the CMS limits presented in \cite{Cms:2024,CMS:2018jxx} are not sensitive.

		\item In the right panel of FIG.~\ref{fig:cmslimit}, the variation of $p p \rightarrow e N \rightarrow eejj$ cross section with $M_N$ is shown for different combinations of $C_{\Lambda}$ and $C_{\Lambda}^{\prime}$. The lines correspond to the same Wilson coefficients mentioned earlier, maintaining the same colour code. The grey shaded region is the upper limit on the $p p \rightarrow e N \rightarrow eejj$ cross section obtained from the CMS search \cite{CMS:2022chw}. As can be seen, a signal with higher $C_{\Lambda}^{\prime}$, marked by the pink line, is heavily constrained up to 3 TeV. For \((C_{\Lambda} = 4 \times 10^{-9}, C_{\Lambda}' = 4 \times 10^{-8})\) and \((C_{\Lambda} = 10^{-8}, C_{\Lambda}' = 10^{-7})\), constraints are present in the mass ranges \(500 < M_N < 1500\) GeV and \(500 < M_N < 3000\) GeV, respectively. For other combinations of Wilson coefficients, the signal remains unconstrained. It's important to note that in this case, there is a one-to-one correspondence between the signal topology investigated in \cite{CMS:2022chw} and the above mentioned figure and hence we expect the cut-efficiencies will also be similar. 
        Therefore, the argument about the influence of cut efficiency, which was relevant in the left plot, doesn't apply here. However, with the increase of $C_{\Lambda}$, the branching $N \to e j j$ will decrease, and the parameter space will remain unconstrained.
	\end{itemize}

 In our subsequent discussion, we adopt a conservative approach and consider two sets of benchmark values of $C_{\Lambda}$ and $C^{\prime}_{\Lambda}$ as $C_{\Lambda} = 7 \times 10^{-8}, C^{\prime}_{\Lambda} = 7 \times 10^{-9} $ and  $C_{\Lambda} = 4 \times 10^{-9}, C^{\prime}_{\Lambda} = 4 \times 10^{-8}$ which are unconstrained in a wider range of $M_N$ in between 10 GeV to 500 GeV from the searches mentioned above \cite{Cms:2024,CMS:2018jxx,CMS:2022chw} and explore the discovery prospect of tri-lepton+MET final state at the HL-LHC. For completeness, we also extend the analysis for $M_N$ between 500 GeV and 1 TeV. Despite the small branching ratio of $N \to e l \nu$ final state, which is $(20-30)\%$ for $C_{\Lambda} > C^{\prime}_{\Lambda}$ and $(0.4-20)\%$ for $C_{\Lambda} < C^{\prime}_{\Lambda}$, the presence of multi-leptons provide a cleaner final state at the HL-LHC and hence may lead to better sensitivity. 
	
\section{Interplay of \texorpdfstring{$\mathcal{O}_{HNe}$ and $\mathcal{O}_{duNe}$}{} in tri-lepton + MET search \label{sec:5}}

Before proceeding further with the analysis of HL-LHC, we discuss  the kinematic variables used in CMS $3\ell + \slashed{E}$ search \cite{CMS:2018iaf} and its implication for heavy neutral lepton search at HL-LHC for four different scenarios: 	

a) $M_N < M_W$, $C_{\Lambda} > C^{\prime}_{\Lambda}$, b) $M_N < M_W$, $C_{\Lambda} < C^{\prime}_{\Lambda}$, c) $M_N > M_W$, $C_{\Lambda} > C^{\prime}_{\Lambda}$, and d)
$M_N > M_W$, $C_{\Lambda} < C^{\prime}_{\Lambda}$.

We closely follow  \cite{CMS:2018iaf} and later propose our own sets of kinematic variables. We notice that the strategy mentioned in the most recent search \cite{Cms:2024} has a similar outcome as \cite{CMS:2018iaf}. We first consider two illustrative mass points, $M_N = 50$ GeV and $M_N = 450$ GeV, to show the distributions of different kinematic variables, evaluate the cut efficiencies, and analyse the signal significance. Subsequently, we vary $M_N$ over a wider range from 10 GeV to 1 TeV. The above-mentioned benchmark mass points are also used for BDT analysis. For each of these two mass points, we set $C_{\Lambda} = 7 \times 10^{-8}$, $C^{\prime}_{\Lambda} = 7 \times 10^{-9}$ $(C_{\Lambda}<C^{\prime}_{\Lambda})$ and  $C_{\Lambda} = 4 \times 10^{-9}$, $C^{\prime}_{\Lambda} = 4 \times 10^{-8}$ $(C_{\Lambda}>C^{\prime}_{\Lambda})$. We consider the c.m. energy 14 TeV and luminosity 3000 $\text{fb}^{-1}$ as expected for HL-LHC. For both signal and background generation, we include $\tau$ as part of the leptons, implying that we consider $W^{\pm} \rightarrow e^{\pm}/\mu^{\pm}/\tau^{\pm} + \nu/\bar{\nu}$ processes with subsequent leptonic decays of $\tau$. Note that, in \cite{CMS:2018iaf} $\ell$ represents both $e$ and $\mu$. Since, in our case $N$ interacts only with $e$, therefore the above mentioned signature ($3\ell + \slashed{E}$) implies $2e + (e / \mu ) + \slashed{E}$. To distinguish the SM background from the signal, the CMS analysis used the following variables: flavour and charge of the selected leptons, number of leptons and number of $b-$jets, $p_T$ of the leptons,   $M_{3\ell}$, $M_{\ell^+ \ell^-}$ and $p_T^{\mathrm{miss}}$.  In the above, $M_{3\ell}$ is the invariant mass of the tri-lepton system. This variable is used only for $M_N < M_W$ scenario, since in this scenario, the $W^\pm$ is produced on-shell and, hence, for signal $M_{3\ell}$ should be less than $M_W$.
$M_{\ell^+ \ell^-}$ is the invariant mass of the opposite sign lepton pair closer to the mass of $Z$-boson, and $p_T^{\mathrm{miss}}$ is the missing transverse momentum of each event. A moderate $p_T^{\mathrm{miss}}$ cut is applied for both high ($M_N > M_W$) and low ($M_N < M_W$) $M_N$.

\begin{table*}[hbt!]
	\centering
	\small 
	\begin{adjustbox}{width=0.85\textwidth} 
		\addtolength{\tabcolsep}{-1pt}
		\begin{tabular}{||c|c|c|c|c|c|c|c||}
			\hline 
			& no. bjets = 0  & no. $\ell^\pm = 3$ &  $M_{3\ell} < 80$ GeV & $|M_{\ell^\pm \ell^\mp} - M_Z| > 15$   GeV  & $p_T^{\mathrm{miss}} < 75$ GeV& ${\sigma}_{\text{eff}}$[fb] &${\eta}_{s}$  \\
			\hline
			50 GeV($C_{\Lambda} > C_{\Lambda}^{\prime}$)\, [6.59 fb]&  6.59 & 1.34 & 1.30 & 1.30 & 1.29&1.29 & {26.86}\\
			50 GeV($C_{\Lambda} < C_{\Lambda}^{\prime}$)\, [2.25 $\times 10^{-2}$ fb] &  2.24 $\times 10^{-2}$ \footnote{The reduction in the effective signal cross-section from 2.25 $\times 10^{-2}$ fb to 2.24 $\times 10^{-2}$ fb occurs due to the imposed constraint $d_l<2$mm.}  & 3.94 $\times 10^{-3}$ & 9.93 $\times 10^{-5}$& 9.93 $\times 10^{-5}$&  9.89 $\times 10^{-5}$& 9.89 $\times 10^{-5}$ & {2.23$\times 10^{-3}$}\\
			\hline \hline
			$VV $  [9847 fb]     & 9847 & 277.29 & 2.19 & 1.70 & 1.22& 1.22 & \\
			$VVV$  [5.53 fb]      & 5.53 & 1.22 & 5.74 $\times 10^{-2}$ & 4.29 $\times 10^{-2}$ & 2.76 $\times 10^{-2}$ & 2.76 $\times 10^{-2}$  &  \\
			$t\bar{t}$  [51170 fb]  & 11289.56 & 92.18 & 11.57 & 8.72 & 4.35 & 4.35 & \\
			$t\bar{t}W^\pm$  [13.22 fb]  & 2.29 & 4.83 $\times 10^{-1}$ & 3.16 $\times 10^{-2}$ & 2.31 $\times 10^{-2}$ & 9.26 $\times 10^{-3}$ & 9.26 $\times 10^{-3}$  &\\				
			$t\bar{t}Z$  [7.26 fb]  & 1.31 & 3.99 $\times 10^{-1}$ & 8.67 $\times 10^{-3}$ & 6.66 $\times 10^{-3}$ & 2.87 $\times 10^{-3}$ & 2.87 $\times 10^{-3}$ &\\
			\hline 
		\end{tabular}
	\end{adjustbox}
	\caption{Cross-sections of signal and backgrounds, after imposing CMS-motivated cuts as outlined in \cite{CMS:2018iaf}. We consider $M_N = 50 $ GeV and $C_{\Lambda} =4 \times 10^{-9}$ $(7 \times 10^{-8})  $, $C_{\Lambda}^{\prime} = 4 \times 10^{-8}$ $( 7 \times 10^{-9})$.}
	\label{tab:CMSlowmass}
\end{table*}

\begin{table*}[hbt!]
	\centering
	\small 
	\begin{adjustbox}{width=0.8\textwidth} 
		\addtolength{\tabcolsep}{-1pt}
		\begin{tabular}{||c|c|c|c|c|c|c||}
			\hline 
			& no. bjets = 0  & no. $\ell^\pm = 3$  & $|M_{\ell^\pm \ell^\mp} - M_Z| > 15$   GeV    & $p_T^{\mathrm{miss}} > 50$ GeV& ${\sigma}_{\text{eff}}$[fb]&${\eta}_{s}$  \\
			\hline
			450 GeV($C_{\Lambda} > C_{\Lambda}^{\prime}$)\, [2.2 $\times 10^{-2}$ fb]&  2.2 $\times 10^{-2}$ & 1.17 $\times 10^{-2}$  & 1.15 $\times 10^{-2}$ & 8.96 $\times 10^{-3}$ &8.96 $\times 10^{-3}$ &{5.7 $\times 10^{-2}$}\\
			450 GeV($C_{\Lambda} < C_{\Lambda}^{\prime}$)\, [7.6 $\times 10^{-2}$ fb] &  7.6 $\times 10^{-2}$ & 4.02 $\times 10^{-2}$  & 3.96 $\times 10^{-2}$  & 3.08 $\times 10^{-2}$ & 3.08 $\times 10^{-2}$ &{1.96 $\times 10^{-1}$}\\
			\hline \hline
			$VV $  [9847 fb]     & 9847 & 202.31 & 115.07 & 50.95& 50.95 &\\
			$VVV$  [5.53 fb]      & 5.53 & 9.27 $\times 10^{-1}$  & 5.53 $\times 10^{-1}$ & 3.84 $\times 10^{-1}$ &3.84 $\times 10^{-1}$&\\
			$t\bar{t}$  [51170 fb]  & 11289.56 & 55.94 & 31.43 & 21.66& 21.66 &\\
			$t\bar{t}W^\pm$  [13.22 fb] & 2.29 & 3.85 $\times 10^{-1}$ & 2.29 $\times 10^{-1}$ & 1.79 $\times 10^{-1}$ & 1.79 $\times 10^{-1}$ &\\				
			$t\bar{t}Z$  [7.26 fb] & 1.31 & 3.50 $\times 10^{-1}$ & 1.79 $\times 10^{-1}$ &1.41 $\times 10^{-1}$ & 1.41 $\times 10^{-1}$&\\			
				\hline 
		\end{tabular}
	\end{adjustbox}
		\caption{Cross-sections of signal and backgrounds, after imposing CMS-motivated cuts mentioned in \cite{CMS:2018iaf}. We consider $M_N = 450 $ GeV, $C_{\Lambda} =4 \times 10^{-9}$ $(7 \times 10^{-8})$, and  $C_{\Lambda}^{\prime} = 4 \times 10^{-8}$ $( 7 \times 10^{-9})$.}
		\label{tab:CMShighmass}
\end{table*}

\begin{figure}[h]
	\centering

	\includegraphics[scale=0.37]{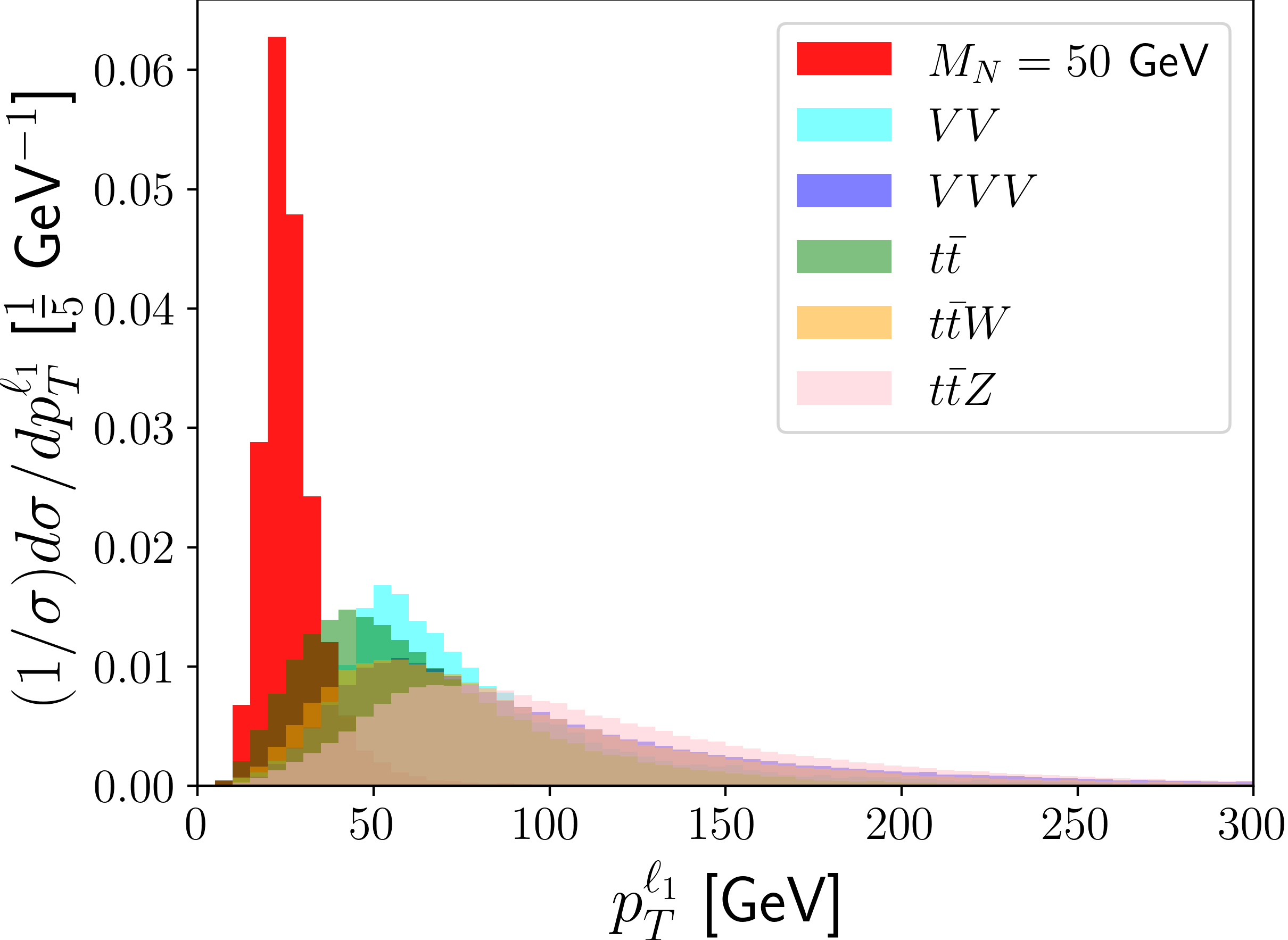}
	\includegraphics[scale=0.37]{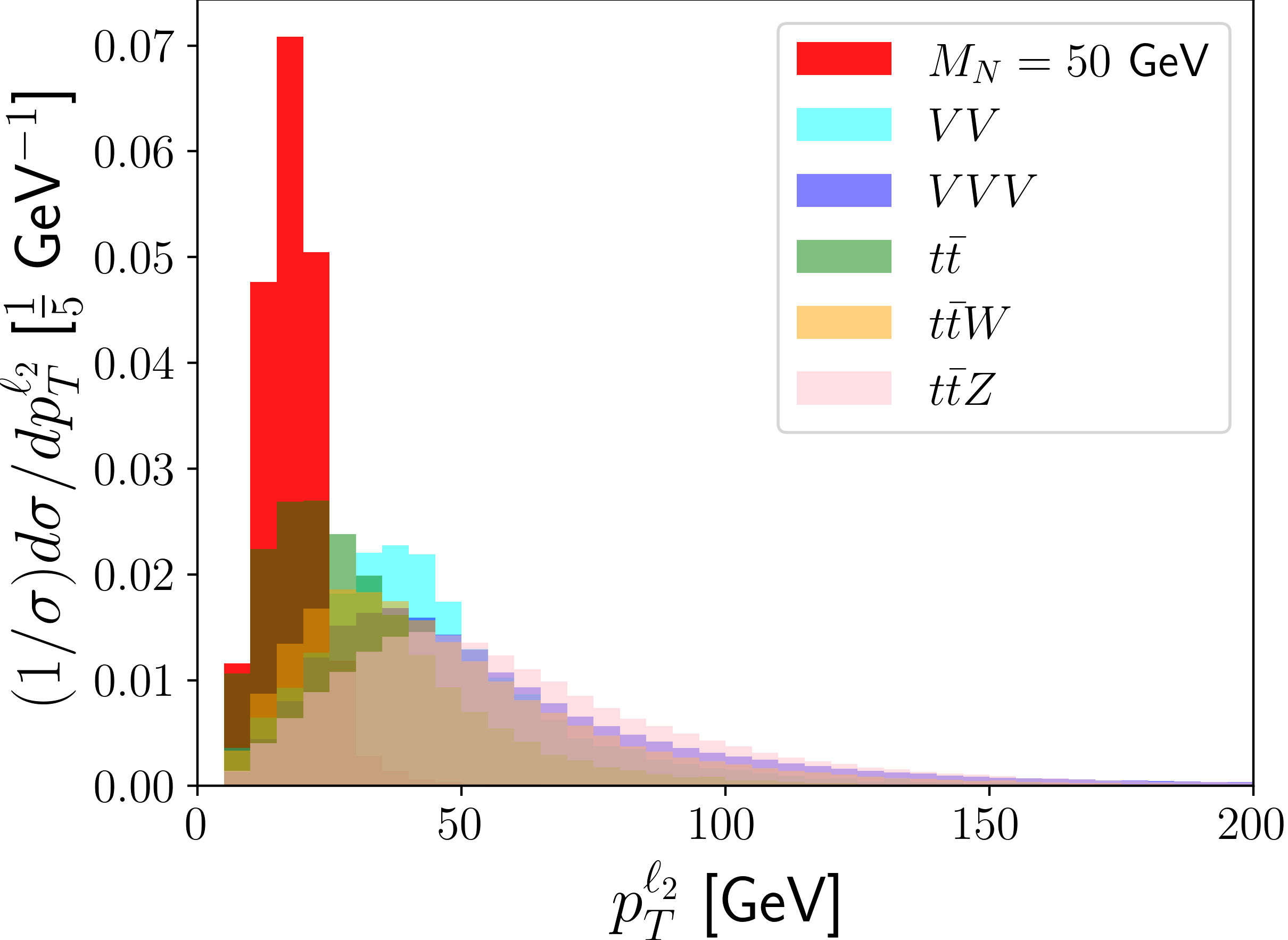}
	\includegraphics[scale=0.37]{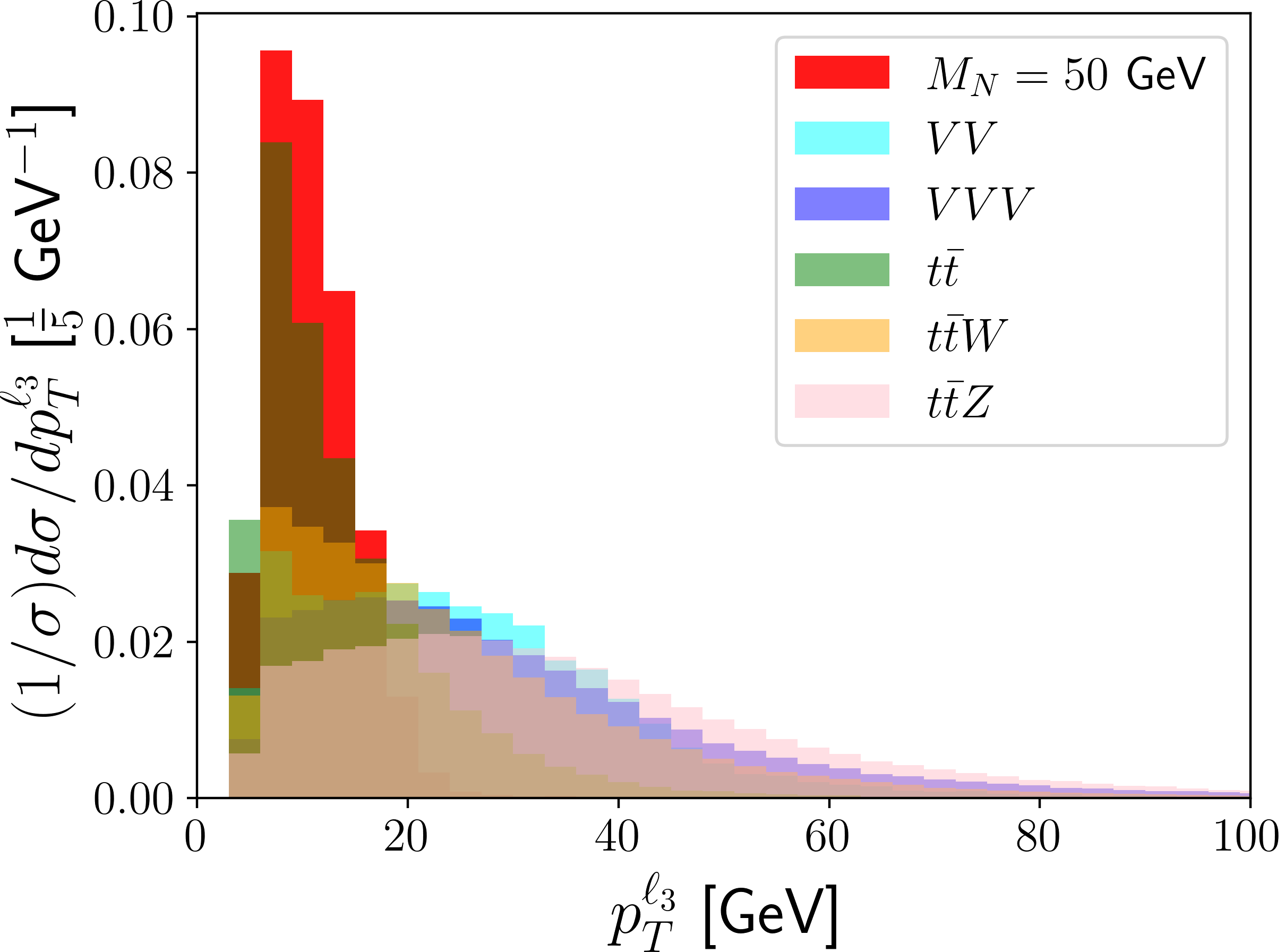}
	\includegraphics[scale=0.37]{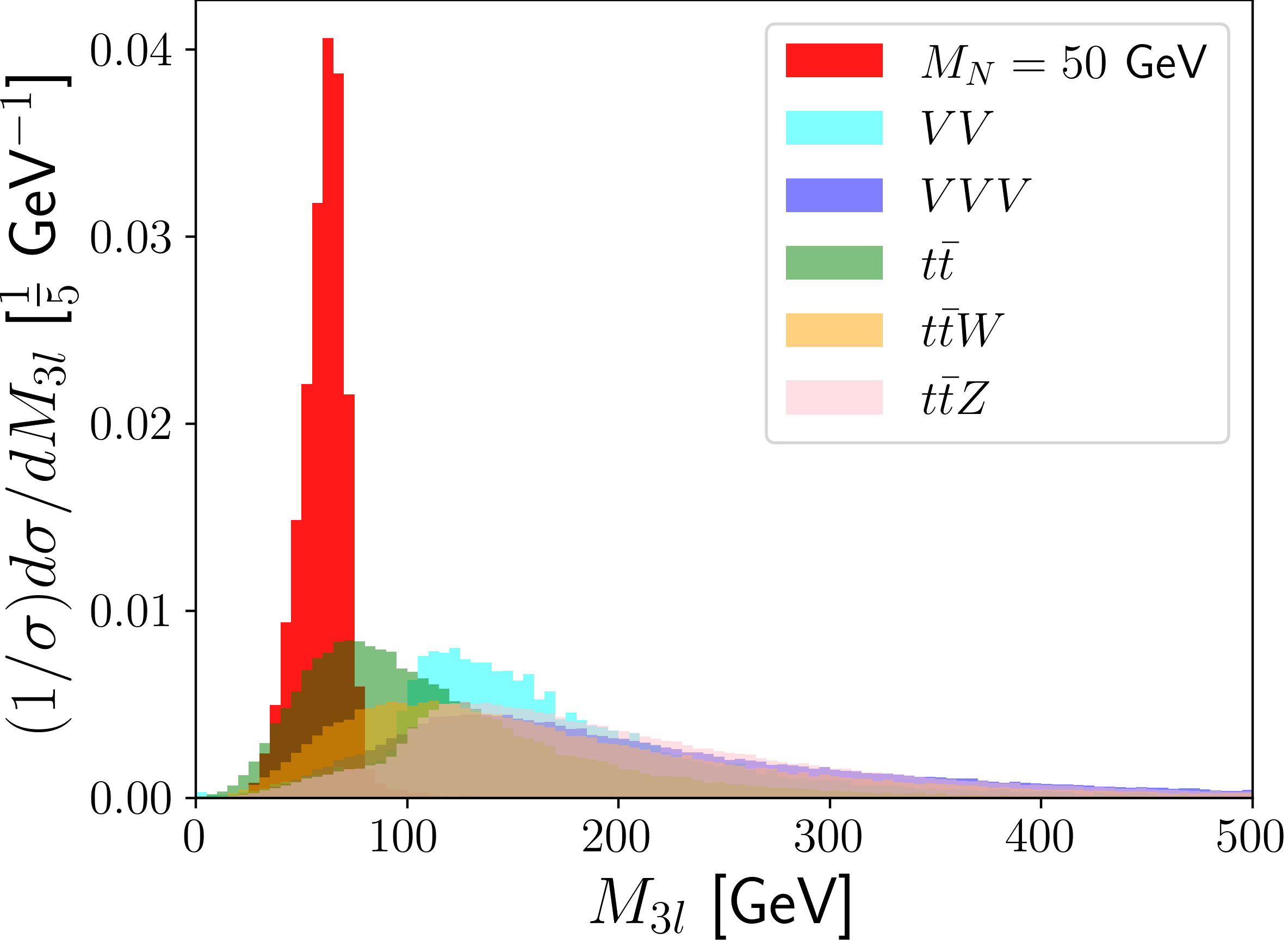}
	\includegraphics[scale=0.37]{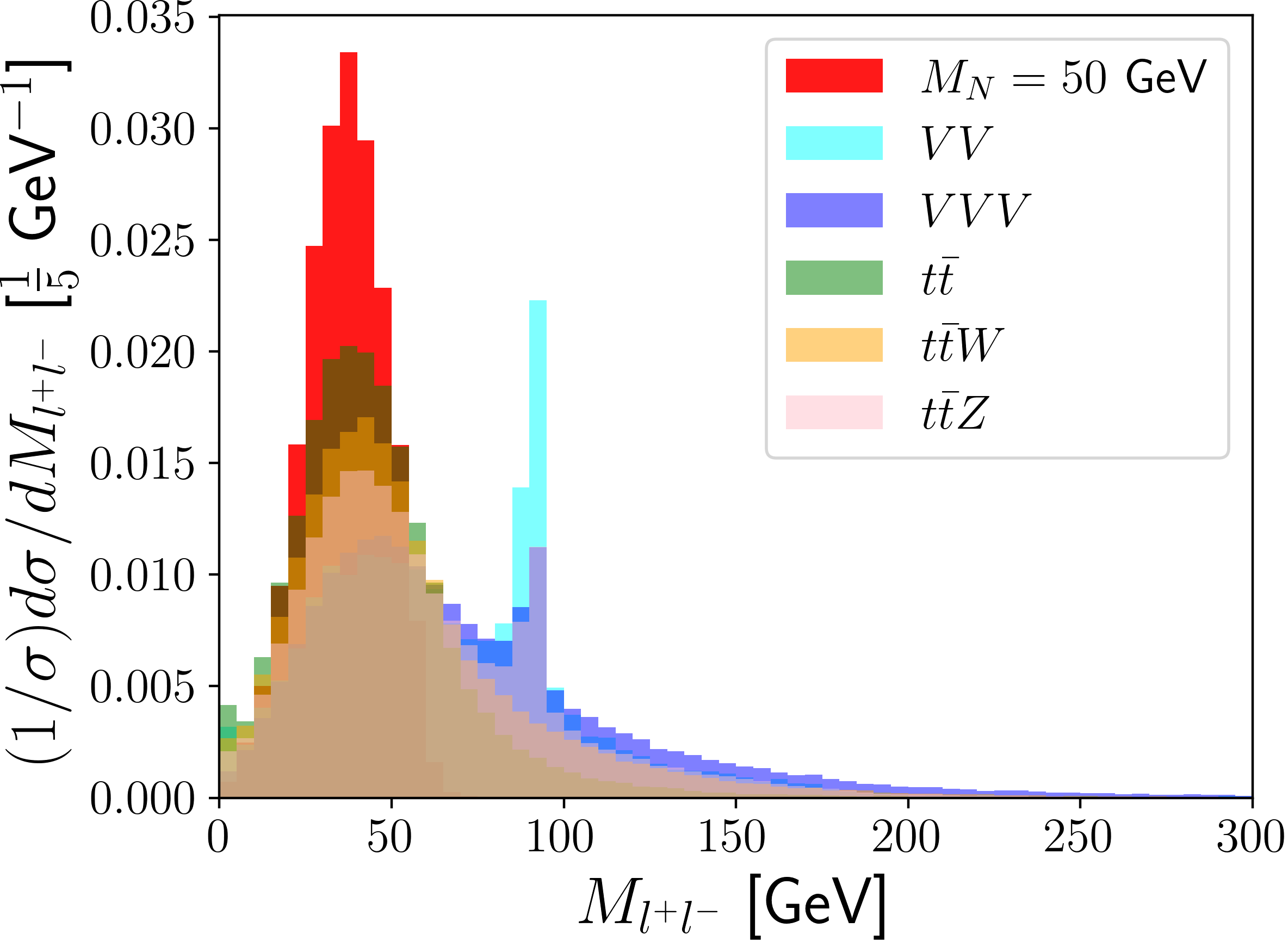}	
	\includegraphics[scale=0.37]{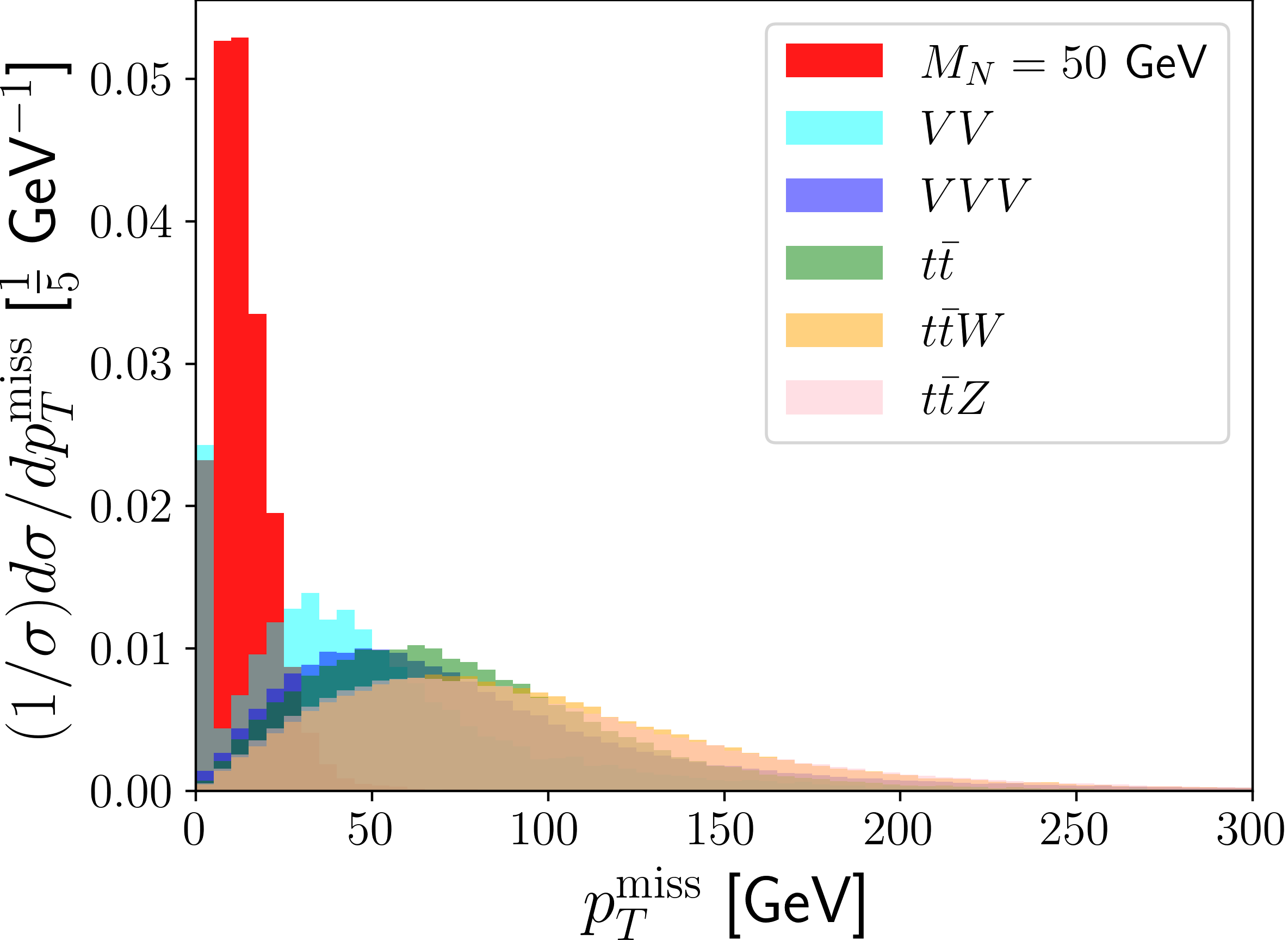}

	\caption{ The figure showcases different kinematic variables which were utilised for $M_N = 50$ GeV, $C_\Lambda = 7 \times 10^{-8}$, and $C^\prime_\Lambda = 7 \times 10^{-9}$.}
	\label{Fig:dist_lowmn_cgtcp}
\end{figure}

	For background, we consider SM processes $VV, VVV (V=W,Z)$, $t \bar{t}$, $t \bar{t}+W/Z$ with their subsequent decays. They were generated in madgraph with $p_T$ cut, $p_T > 10$ GeV on the final state leptons. We generate $VV+j$ merged sample, for which we consider ${xq}_{cut} = 30$ GeV and $q_{cut} = 45$ GeV respectively.   We show the distributions of different kinematic variables, both for the signal and backgrounds, in  FIG.~\ref{Fig:dist_lowmn_cgtcp}, FIG.~\ref{Fig:dist_lowmn_cpgtc}, and  FIG.~\ref{Fig:dist_highmn_cpgtc}.

Motivated by the analysis presented in \cite{CMS:2018iaf}, we use a somewhat simplistic set of selection cuts, see TABLE~[\ref{tab:CMSlowmass},\ref{tab:CMShighmass}], to distinguish between the signal and the SM background. As noted in \cite{CMS:2018iaf}, we implement a loose isolation criterion for all the selected leptons, defined as $I_{\text{rel}} < 0.6$. Here, $I_{\text{rel}}$ represents the ratio of the scalar sum of $p_T$ of particles within a specified cone size (0.3) around the lepton to the lepton's own $p_T$. We enforce the lepton selection criteria to be $p_T >10 $ GeV and $|\eta^\ell| > 2.5$, both for $e$ and $\mu$. We checked that by employing a dynamic isolation strategy based on the transverse momentum of the lepton, as mentioned in \cite{Cms:2024}, we achieve similar cut efficiency. For $b$-jet tagging, we designate jets within a 0.4 radius of the $b$-quark as $b$-jets with 85$\%$ tagging efficiency. 
 For the signal, we consider the event samples for which $N$ decays promptly so that the CMS search \cite{CMS:2018iaf} strategy can be followed. For a decay length of $N$ larger than the mm range, the displaced decay search, such as \cite{CMS:2022fut}, is more appropriate, which we do not consider in this work.   We ensure this by  constraining the decay length of RHN $d_{l} < 2$ mm \cite{CMS:2022fut}, which for $M_N >35$ GeV is automatically satisfied for our chosen $C_{\Lambda}$ and $C^{\prime}_{\Lambda}$. Note that, for our signal, requiring a $no.~\ell^\pm = 3$ cut in both TABLE~\ref{tab:CMSlowmass} and TABLE~\ref{tab:CMShighmass} implies having at least two electrons.

For TABLE~[\ref{tab:CMSlowmass}] which corresponds to $M_N=50 $ GeV, and for the scenario $C_{\Lambda}> C^{\prime}_{\Lambda}$, 
we obtain large significance, ${\eta}_{s} = 26.86 \sigma$ following CMS-inspired cuts \cite{CMS:2018iaf}, where we have followed Eq.~\ref{eq:significance} to obtain significance. This primarily occurs due to higher signal-cross section $\sigma(p p \to e N \to e e l \nu)_{\sqrt{s}=14 \rm{TeV}}=6.59$ fb, as well as due to moderate cut-efficiency {  $c_{eff} \sim 0.2$. However, the same set of cuts performs poorly and leads to a smaller cut-efficiency $c_{eff} \sim 4.0\times 10^{-3}$ when  $C_\Lambda <C^{\prime}_\Lambda$}. This results in a much smaller significance,  as seen from TABLE.~\ref{tab:CMSlowmass}. The significantly decreased signal cut efficiency is primarily attributed to the isolation requirement $I_{rel} < 0.6$ and $M_{3l} < 80$ GeV. Given that most signal events cluster at higher $M_{3l}$ values, as can be seen from FIG.~\ref{Fig:dist_lowmn_cpgtc}, the number of events passing this cut is notably low. In this case, the partonic cross-section is also smaller $\sigma =0.022$ fb, which is also responsible for obtaining a smaller significance.  

 For  higher mass scenario $M_N=450 $ GeV, as described in TABLE~[\ref{tab:CMShighmass}], the signal cut-efficiency is approximately $c_{eff} \sim 0.4$ for both of the $C_{\Lambda}$ and $C_{\Lambda}^{\prime}$ combinations. However, these sets of cuts are not very effective in suppressing the SM background for $C_{\Lambda} > C_{\Lambda}^{\prime}$ case. Due to the poor performance of the selected cuts, as well as the lower value of the signal cross-section, the significance in this case is ${\eta}_{s}=0.057 \sigma$. Similar conclusion holds for $C_{\Lambda}<C^{\prime}_{\Lambda}$ scenario.  

The preceding discussion reveals that the performance of the presented analysis could be more optimal, except in the scenario where $M_N = 50$ GeV and $C_{\Lambda} > C_{\Lambda}^{\prime}$. So, the implementation of the cuts must be re-evaluated to accommodate these variations. In the following subsections, we reanalyse different discriminating variables and propose using new variables to achieve better significance. We also perform a BDT analysis for $M_N < M_W$, $C_{\Lambda} < C_{\Lambda}^{\prime}$ and $M_N > M_W$, $C_{\Lambda} > C_{\Lambda}^{\prime}$ scenarios, which will be presented in the latter subsection.

\subsection{Redefinition of cuts}
The scenario dictated as a) in the previous section has better significance using CMS-motivated cuts. So, we reanalyse various discriminating variables for the rest of the three scenarios dictated as b), c) and d), where the analysis strategy mentioned in the previous section performs relatively poorly. To do so,  we propose our variables and cuts to better discriminate between signal and SM background. As stated earlier, our focus primarily revolves around $N$, which exclusively interacts with an electron. Consequently, we define our signal as comprising a minimum of three leptons, with a requirement that at least two of them must be $e^\pm$, i.e. $N_{\ell^\pm} \geq 3$ and  $N_{e^\pm} \geq 2$, while the other leptons can be either $e/ \mu$. We also implemented jet veto for $b-$jets. In the subsequent subsections, we will individually discuss scenarios b), c), and d).

\subsubsection{$M_N < M_W$ with $C_{\Lambda}<C^{\prime}_{\Lambda}$  - scenario- b :\label{Subsec:case2}}
\begin{figure}[h]
	\centering

	\includegraphics[scale=0.37]{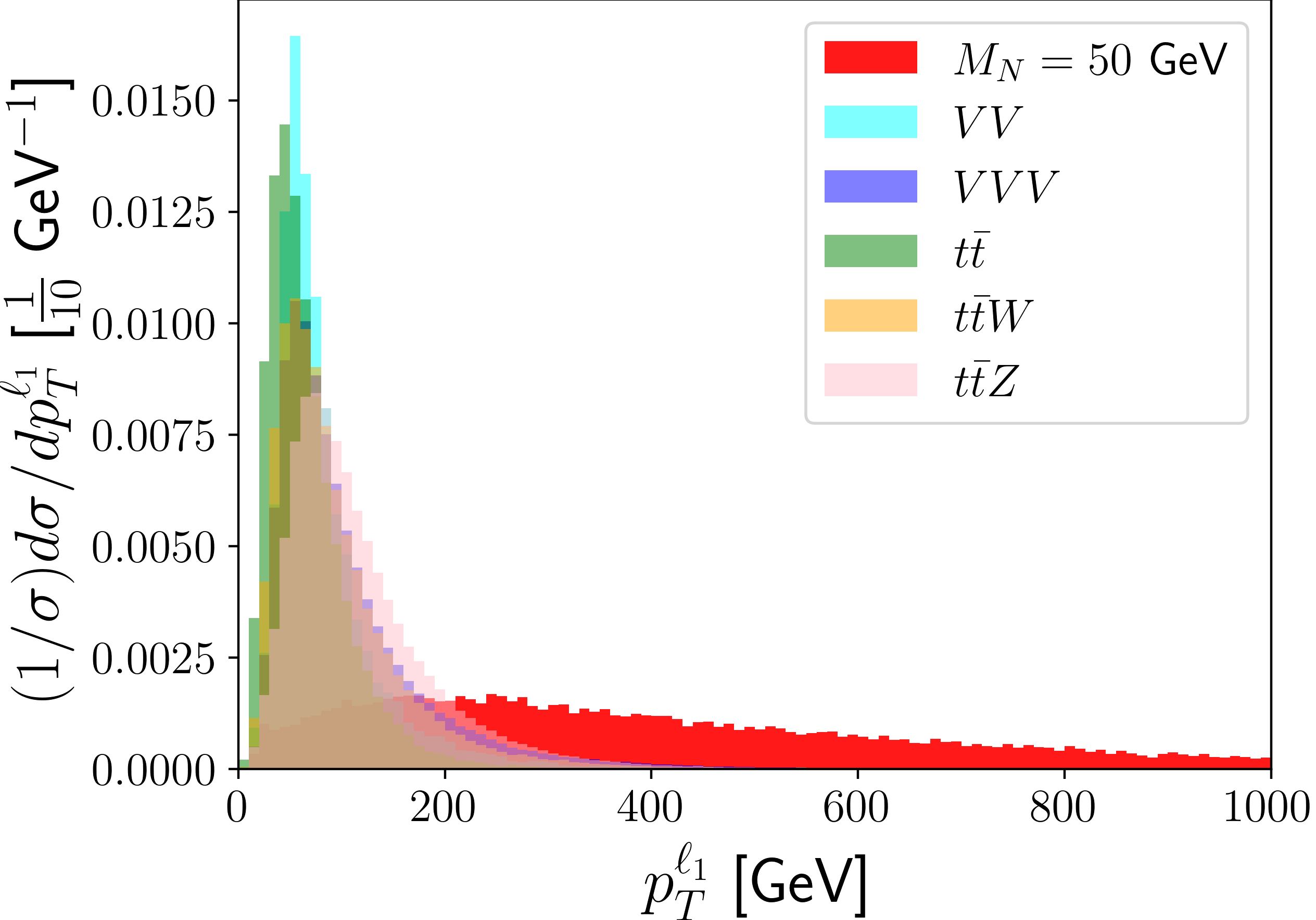}
	\includegraphics[scale=0.37]{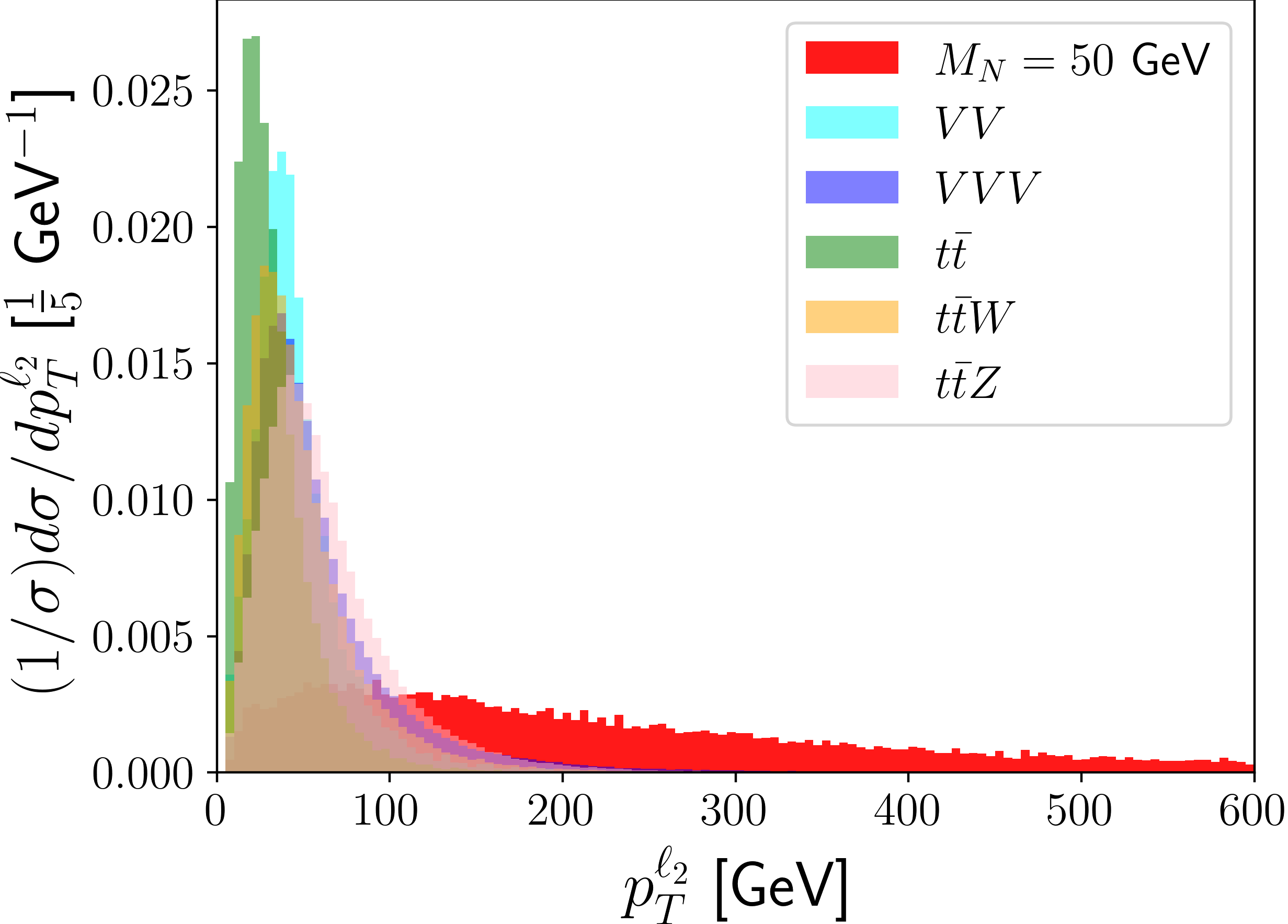}
	\includegraphics[scale=0.37]{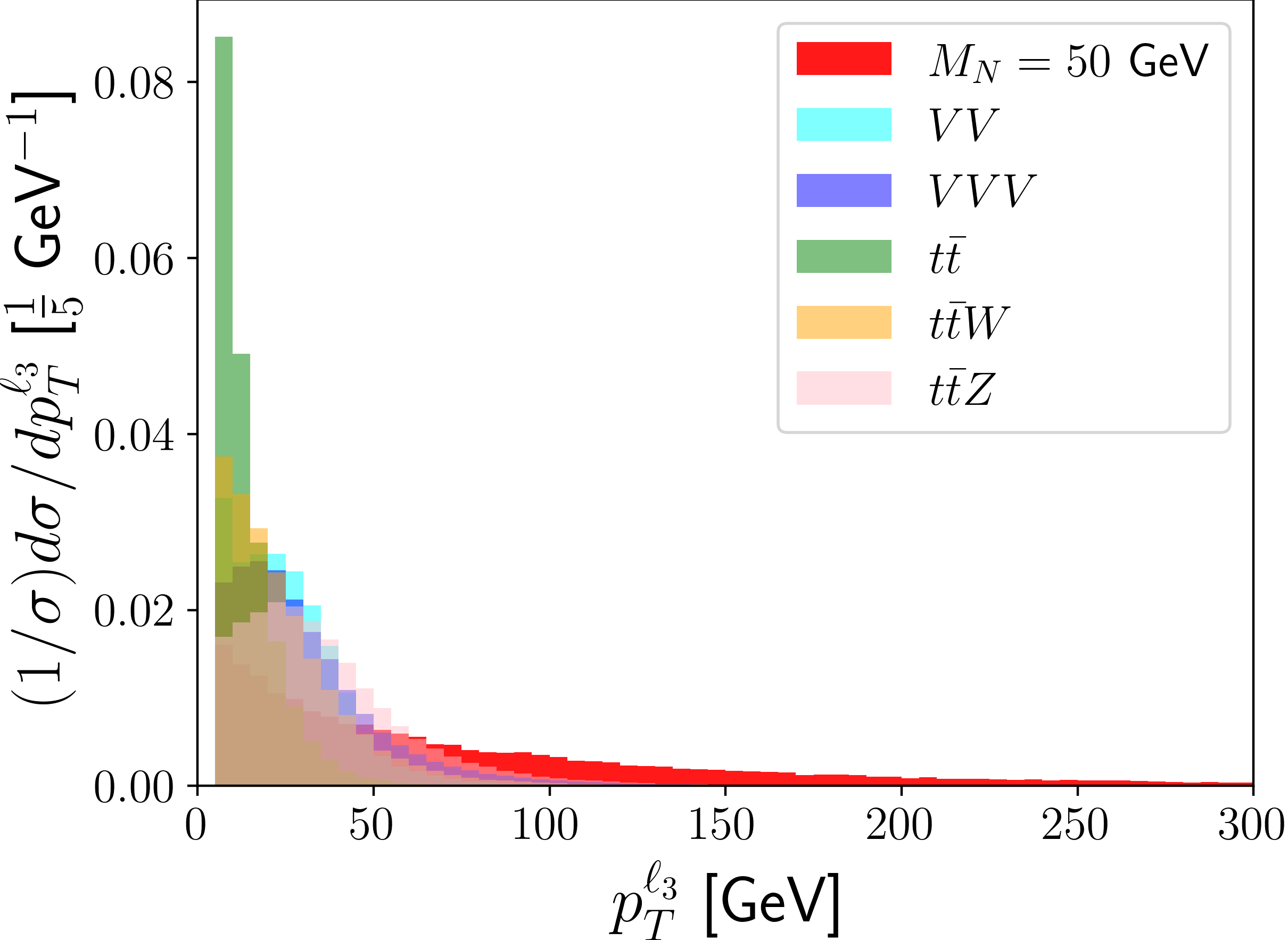}
	\includegraphics[scale=0.37]{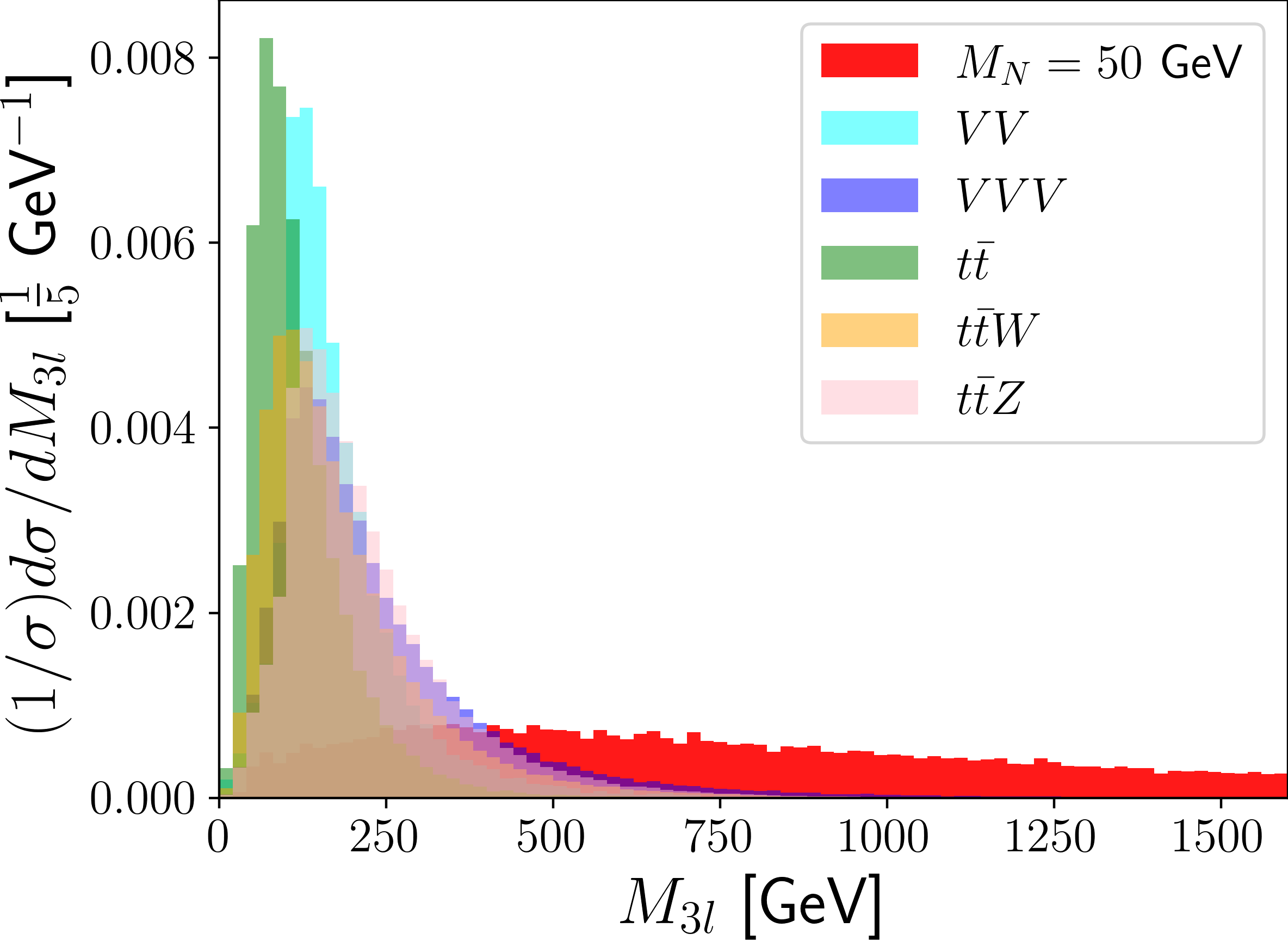}
	\includegraphics[scale=0.37]{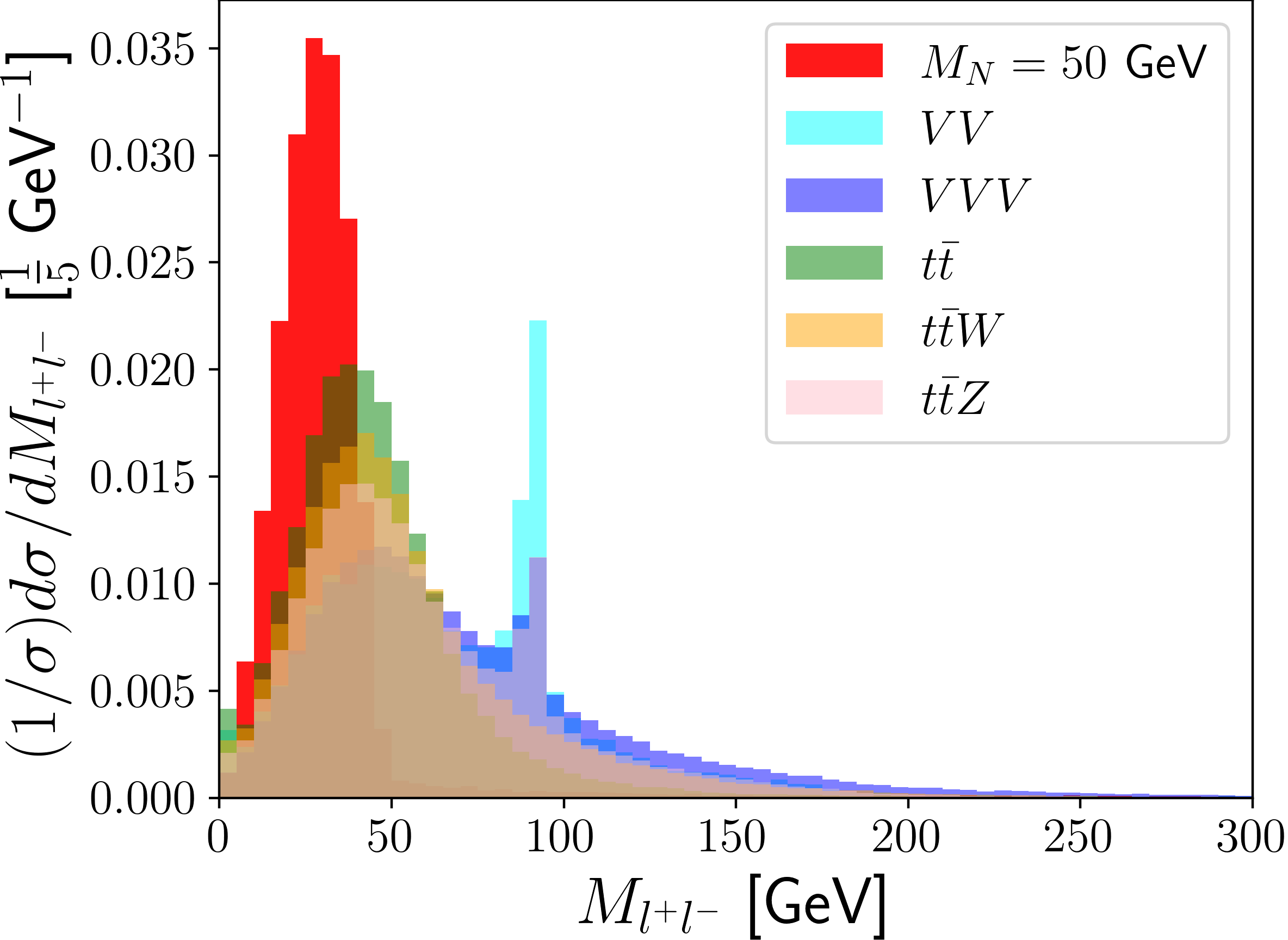}	
	\includegraphics[scale=0.37]{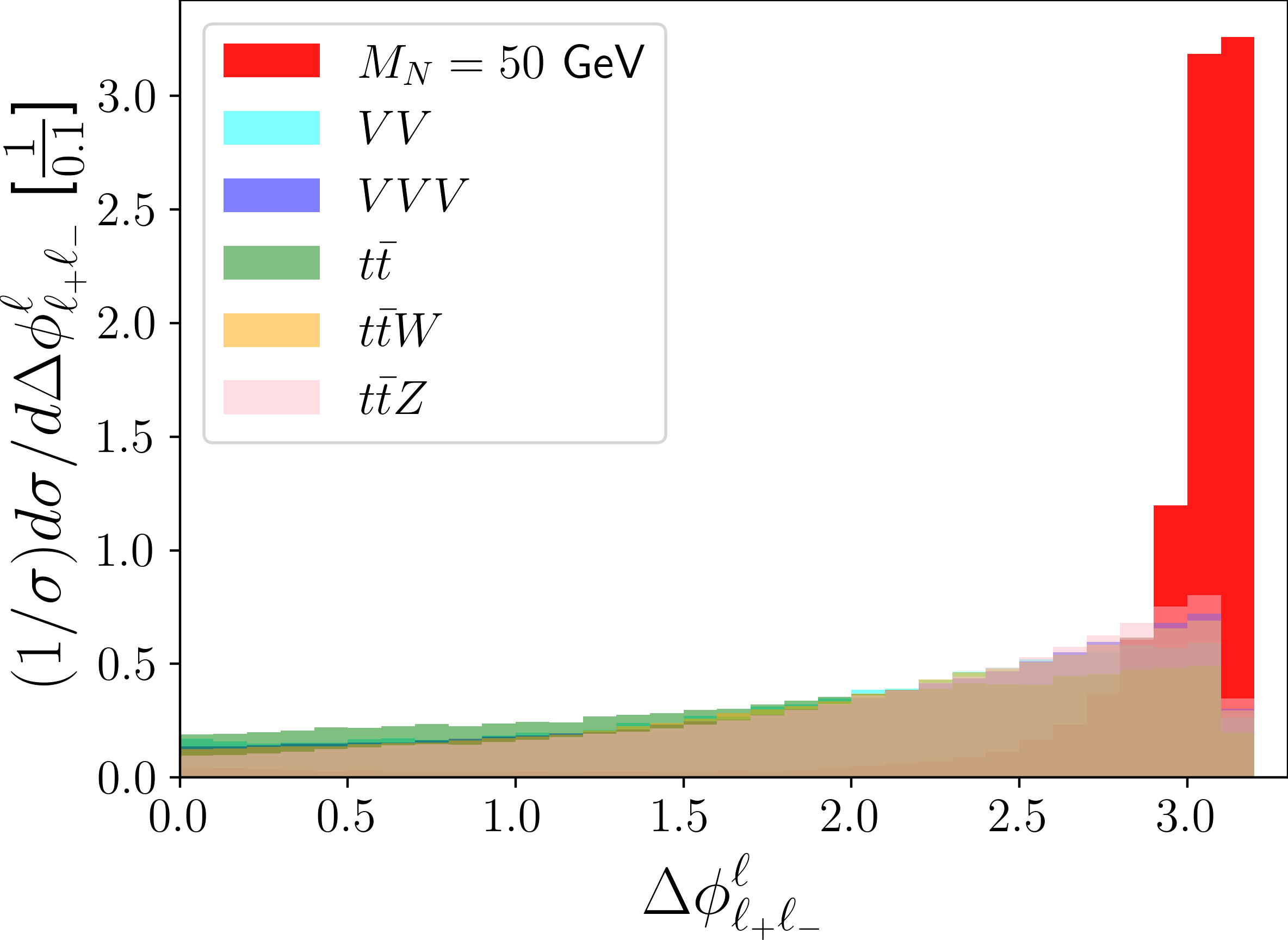}
	\includegraphics[scale=0.37]{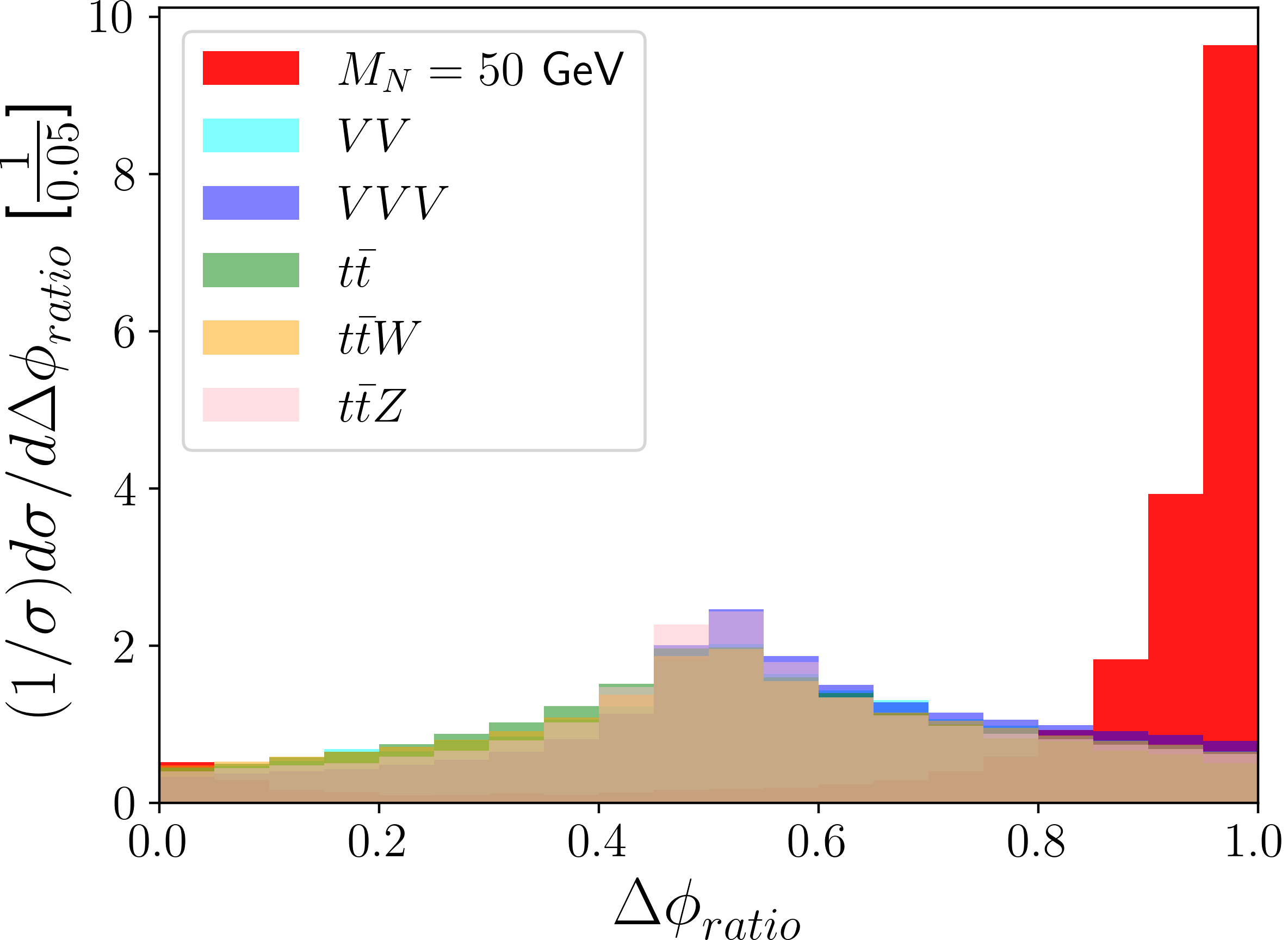}
	\includegraphics[scale=0.37]{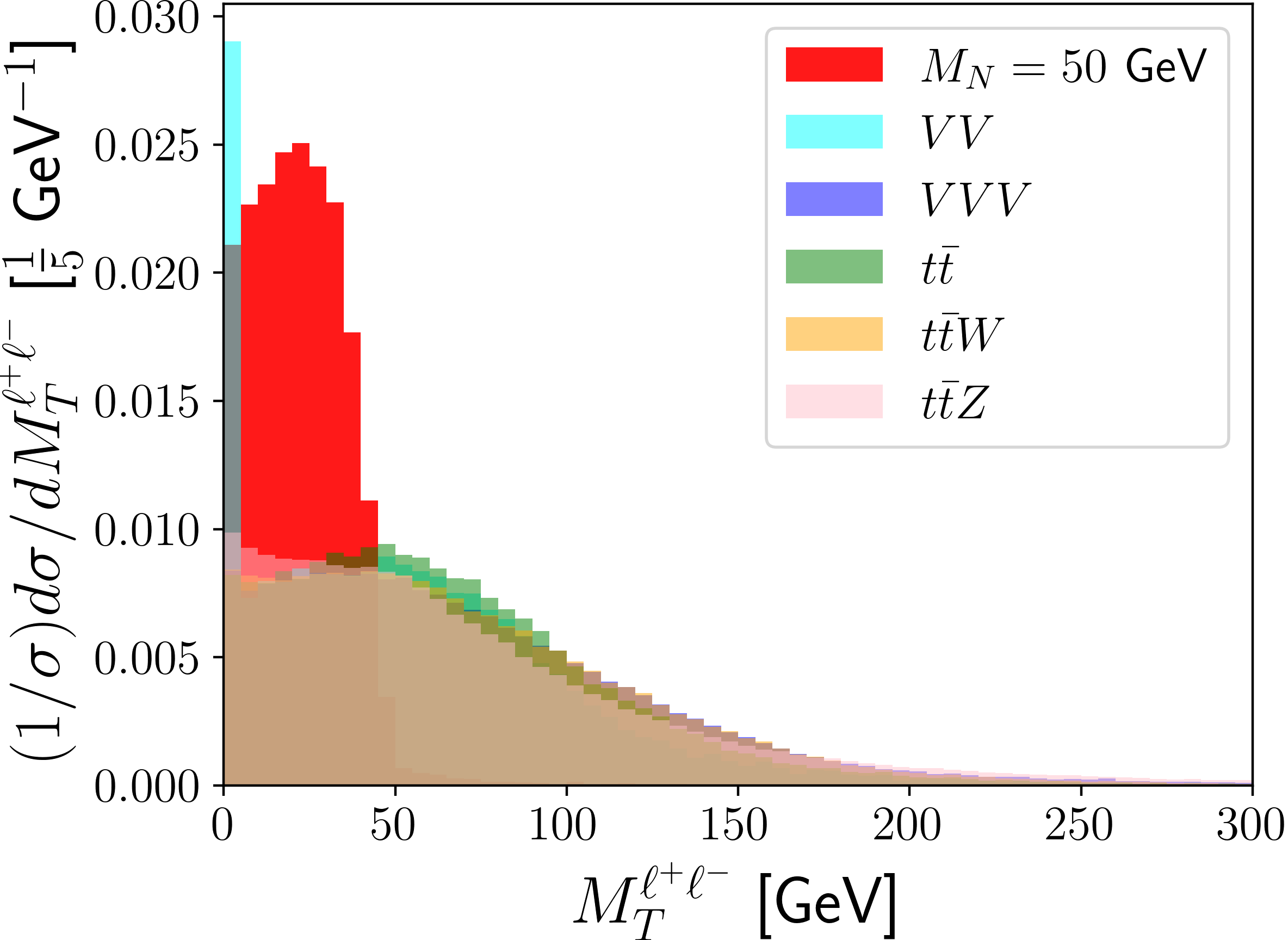}
	\includegraphics[scale=0.37]{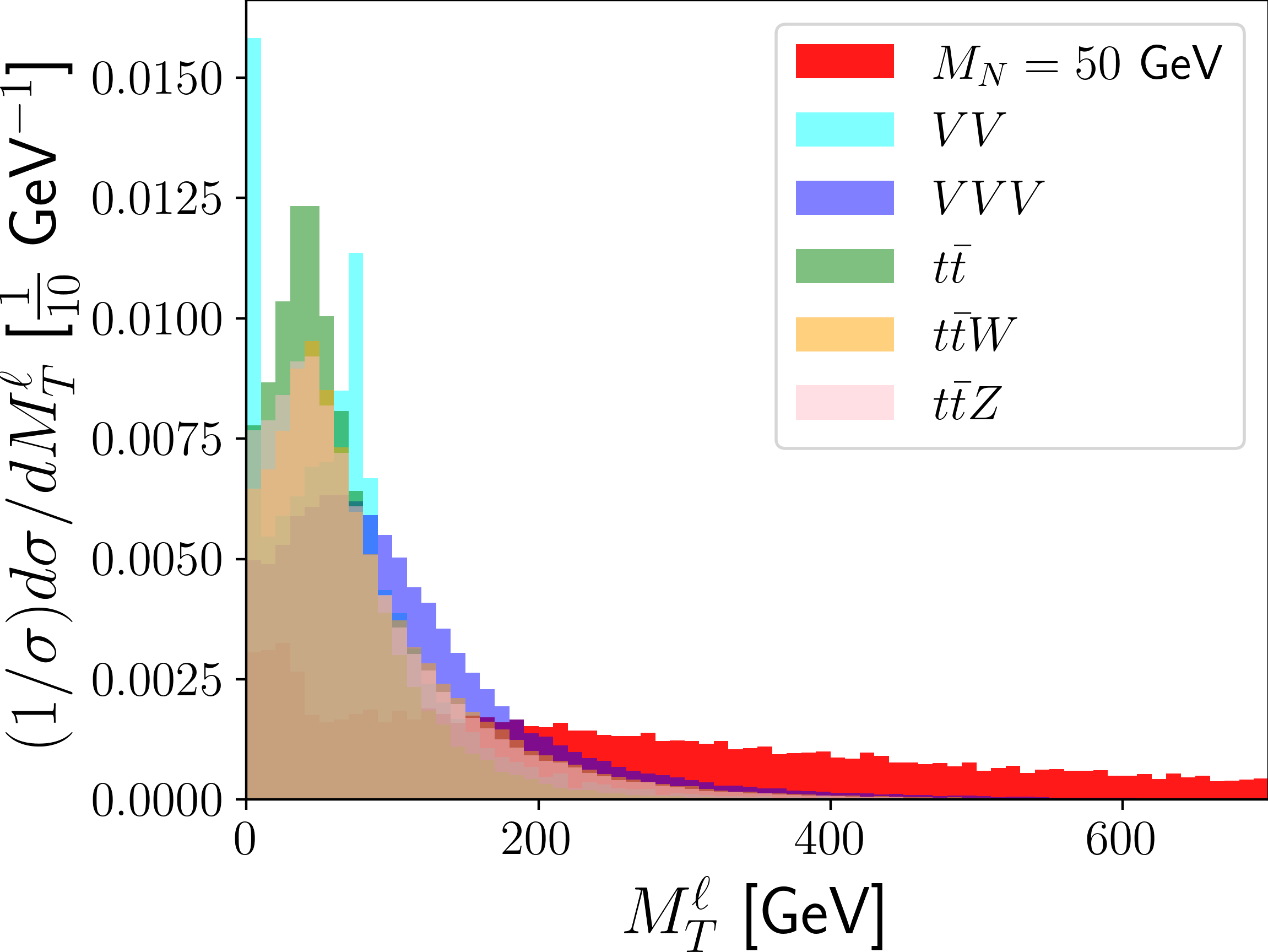}
	
	\caption{ The figure showcases different kinematic variables which were utilised for$M_N = 50$ GeV, $C_\Lambda = 4 \times 10^{-9}$, and $C^\prime_\Lambda = 4 \times 10^{-8}$.}
	\label{Fig:dist_lowmn_cpgtc}
\end{figure}

In FIG.~\ref{Fig:dist_lowmn_cpgtc}, we present the distributions of the variables used to analyse scenario b): $M_N = 50$ GeV $< M_{W}$ and $C_\Lambda < C^\prime_\Lambda$. The $p_T$ of the leading and subleading leptons peaks around 250 and 100 GeV, respectively, with a long tail extending up to several hundred GeV and even reaching a TeV. Along with the criteria mentioned at the beginning of the current section, we hence demand the $p_T$ requirement of the leading two leptons are $p_T^{\ell_1} > 220$ GeV and $p_T^{\ell_2} > 80$ GeV as selection cuts, which helps in minimising the background. Due to the high boost of the produced RHN, the decay products of $N$ exhibit relatively lower isolation. Consequently, enforcing isolation criteria for the leptons leads to a significant decrease in signal events after requiring at least three leptons. So, to enhance the signal significance, we followed  \cite{CMS:2017ucf} and did not demand any lepton isolation. The main distinguishing variables in this case are $M_{3\ell}, M_{\ell^+\ell^-}, M_T^{\ell^+ \ell^-}, M_T^{\ell}, \Delta\phi^{\ell}_{\ell_{+} \ell_{-}}, \Delta\phi_{\mathrm{ratio}}$. $M_T^{\ell^+ \ell^-}$ is the transverse mass of the opposite sign di-lepton system, which is closest to $M_N$, and $M_T^{\ell}$ is the transverse mass of the lepton closer to the $W$ boson mass, which is not included in the opposite sign di-lepton system. The variable $M_{l^+l^-}$ has the same definition as described before.  $M_T^{\ell^+ \ell^-}$ is an important variable for our signal as for this case the RHN decay follows, $N \rightarrow e^{\pm} \ell^{\mp} \nu$, and the $M_T^{\ell^+ \ell^-}$ distribution has a sharp fall at $M_T^{\ell^+ \ell^-} = M_N$. $M_T^{\ell}$ is mainly used to remove the SM background resulting from the leptonic decay of $W$ boson. $\Delta\phi^{\ell}_{\ell_{+} \ell_{-}} $ is defined as the maximum azimuthal angle difference between the aforementioned opposite sign di-lepton system and the rest of the selected lepton. $\Delta\phi^{\ell}_{\ell_{+} \ell_{-}} $ is a very good variable to distinguish the signal from the background. In this scenario, the $C^\prime_\Lambda$ mediated diagram dominates, and the lepton and $N$ are mostly produced back to back, which results in a large $\Delta\phi$ separation between the di-lepton system and the other signal lepton. $\Delta\phi_{\mathrm{ratio}}$ is defined as the $\Delta\phi_{\ell_1}^{p_T^{\mathrm{miss}}}/(\Delta\phi_{\ell_1}^{p_T^{\mathrm{miss}}}+\Delta\phi_{\ell_2}^{p_T^{\mathrm{miss}}})$, where $\ell_{1,2}$ are the leading and sub-leading leptons, respectively and $\Delta\phi_{\ell_{1(2)}}^{p_T^{\mathrm{miss}}}$ represents the angular separation between missing transverse momentum and the leading (subleading) lepton. The probability of selecting the lepton produced in association with \(\nu\) as the leading lepton is small, originating from a $50$ GeV $N$ decay. Consequently, when the lepton directly produced from four-Fermi interaction is selected as the leading one, \(\Delta\phi_{\ell_1}^{p_T^{\mathrm{miss}}}\) is much higher, resulting in a higher value of \(\Delta\phi_{\mathrm{ratio}}\) for most of the signal events. After scrutinizing the distributions as depicted in FIG.~\ref{Fig:dist_lowmn_cpgtc}, we impose cuts on the discriminating variables to maximize the signal-to-background ratio, listed as:  $M_{3\ell} > 550$ GeV, $\Delta\phi_{\mathrm{ratio}} > 0.8$, $\Delta\phi^{\ell}_{\ell_{+} \ell_{-}} > 2.5$, $M_T^{\ell_1} > 5.0$ GeV, $M_T^{\ell^+ \ell^-} < 55.0$ GeV, and $|M_{\ell^\pm \ell^\mp} - M_{Z}| > 50$ GeV.	

\begin{table*}[hbt!]
	\centering
	\small 
	\begin{adjustbox}{width=1\textwidth} 
		\addtolength{\tabcolsep}{-1pt} 
		\begin{tabular}{||c|c|c|c|c|c|c|c|c|c|c||}
			\hline 
			& no.~bjets = 0 & no. $\ell^\pm \geq 3$  & $M_{3\ell} > 550$ & ${\Delta \phi}_{\mathrm{ratio}} > 0.8$ & $\Delta\phi^{\ell}_{\ell_{+} \ell_{-}}  > 2.5$ & $M_T^{\ell} > 5.0$ & $M_T^{\ell^{+}\ell^{-}} < 55.0$ & $|M_{\ell^\pm \ell^\mp} - M_{Z}| > 50$  & ${\sigma}_{\text{eff}}$[fb]& ${\eta}_{s}$ \\
			\hline 
			50 GeV ($C_{\Lambda} < C_{\Lambda}^{\prime}$)  [2.23 $\times 10^{-2}$ fb] & 2.22 $\times 10^{-2}$ \footnote{The reduction in the effective signal cross-section from 2.23 $\times 10^{-2}$ fb to  2.22 $\times 10^{-2}$ fb occurs due to  the imposed constraint $d_l<2$mm.} & 9.25 $\times 10^{-3}$ & 8.37 $\times 10^{-3}$ & 7.73 $\times 10^{-3}$ & 7.57 $\times 10^{-3}$ & 6.73 $\times 10^{-3}$ & 6.71 $\times 10^{-3}$ & 6.34 $\times 10^{-3}$ & 6.34 $\times 10^{-3}$ & {7.78 $\times 10^{-1}$}\\
			\hline 
			\hline
			$VV$ [9847 fb] & 9847 & 3.95 & 1.19 & 4.31 $\times 10^{-1}$ &  2.40 $\times 10^{-1}$ & 2.04 $\times 10^{-1}$& 1.35 $\times 10^{-1}$ &9.89 $\times 10^{-2}$ & 9.89 $\times 10^{-2}$  &\\
			$VVV$  [5.53 fb]& 5.53 & 5.87 $\times 10^{-2}$ & 3.51 $\times 10^{-2}$ & 1.48 $\times 10^{-2}$ & 1.23 $\times 10^{-2}$ & 1.18 $\times 10^{-2}$ & 6.61 $\times 10^{-3}$ & 4.21 $\times 10^{-3}$ & 4.21 $\times 10^{-3}$ &\\
			$t\bar{t}$ [51170 fb] & 11289.56 & 5.78 & 1.09 & 4.35 $\times 10^{-1}$ & 2.30 $\times 10^{-1}$ & 1.84 $\times 10^{-1}$ & 1.23 $\times 10^{-1}$ & 9.21 $\times 10^{-2}$  & 9.21 $\times 10^{-2}$ &\\
			$t\bar{t}W^\pm$  [13.22 fb] & 2.29 & 1.46 $\times 10^{-2}$ & 6.24 $\times 10^{-3}$ & 2.27 $\times 10^{-3}$ & 1.88 $\times 10^{-3}$ & 1.77 $\times 10^{-3}$ & 1.03 $\times 10^{-3}$ & 6.35 $\times 10^{-4}$  & 6.35 $\times 10^{-4}$ &\\
			$t\bar{t}Z$  [7.26 fb] & 1.31 & 3.22 $\times 10^{-2}$ & 9.13 $\times 10^{-3}$ & 3.24 $\times 10^{-3}$ & 2.06 $\times 10^{-3}$& 1.94 $\times 10^{-3}$& 1.29 $\times 10^{-3}$ & 5.08 $\times 10^{-4}$  & 5.08 $\times 10^{-4}$ &\\
			\hline 
		\end{tabular}
	\end{adjustbox}
	\caption{Cross-sections of signal and backgrounds, after imposing our proposed sets of cuts for benchmark point $M_N = 50 $ GeV and $C_{\Lambda} = 4 \times 10^{-9}  $, $C_{\Lambda}^{\prime} = 4 \times  10^{-8}$. }
	\label{tab:cut_flow_mnlwmw_redi}
\end{table*}

 For demonstration purposes, the final cut-flow table is presented in TABLE~\ref{tab:cut_flow_mnlwmw_redi}. We re-iterate that in the third column, when we demanded $no.~\ell^\pm \geq 3$, this implies $N_{\ell^\pm} \geq 3$ and $N_{e^\pm} \geq 2$, along with the previously specified $p_T$ requirements on $\ell$. We find that, due to the large cross-section, even after imposing new cuts, the $VV$ and $t\bar{t}$ both processes still contribute significantly to the background with cross-section after cuts being around 0.09 fb. In comparison, the signal cross-section after cuts is very small, 0.006~fb. The cross-sections of other backgrounds, such as $ VVV, t \bar{t}W, t \bar{t} Z$ are extremely small after cuts. Consequently, the significance remains less than 1$\sigma$. For this scenario, we later pursue a BDT analysis.

\subsubsection{$M_N > M_W$ with $C_{\Lambda}> C^{\prime}_{\Lambda}$  and $C_{\Lambda}< C^{\prime}_{\Lambda}$ - scenarios c and  d \label{Subsec:case1}}
\begin{figure}[htb!]
	\centering

	\includegraphics[scale=0.37]{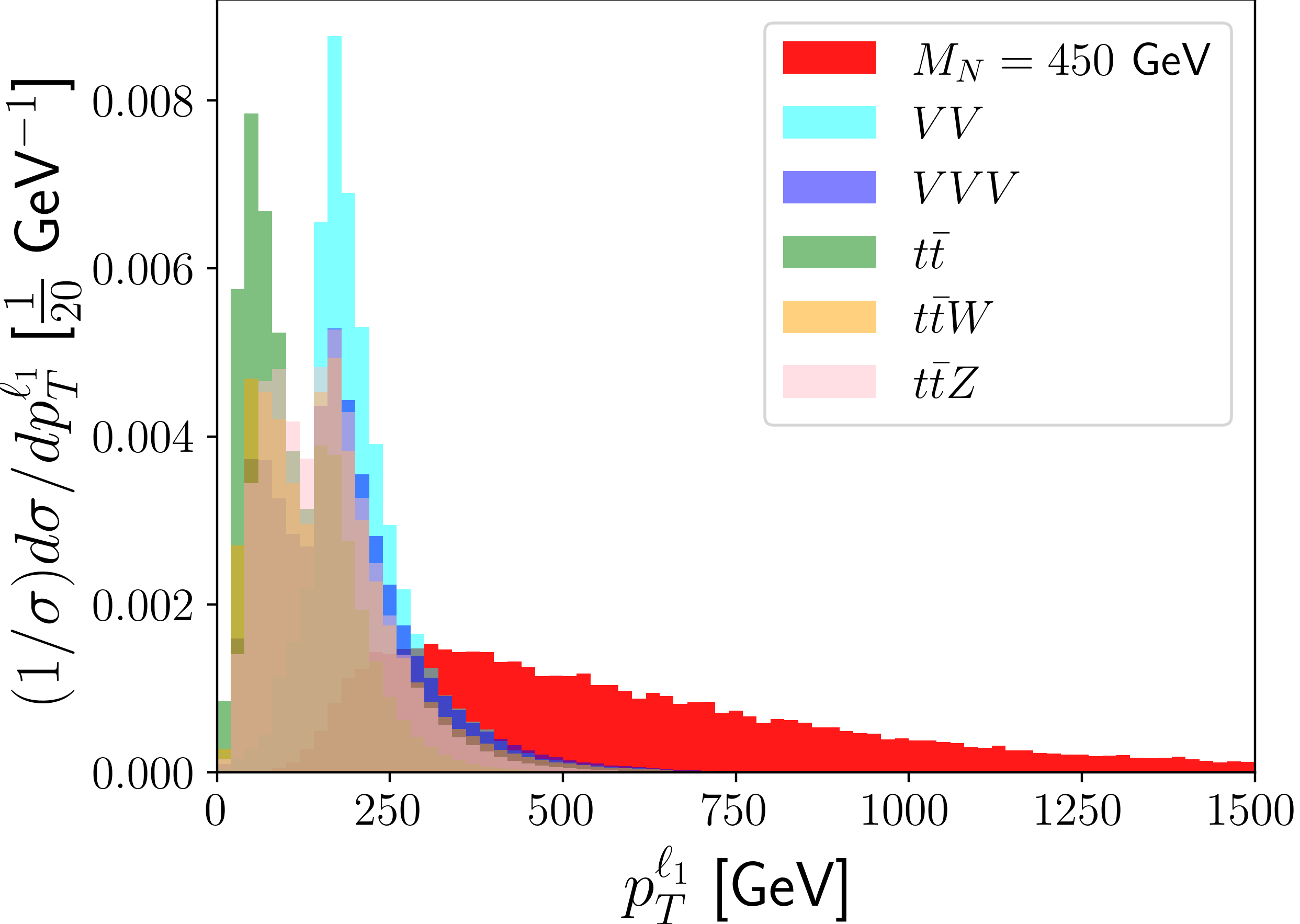}	
	\includegraphics[scale=0.37]{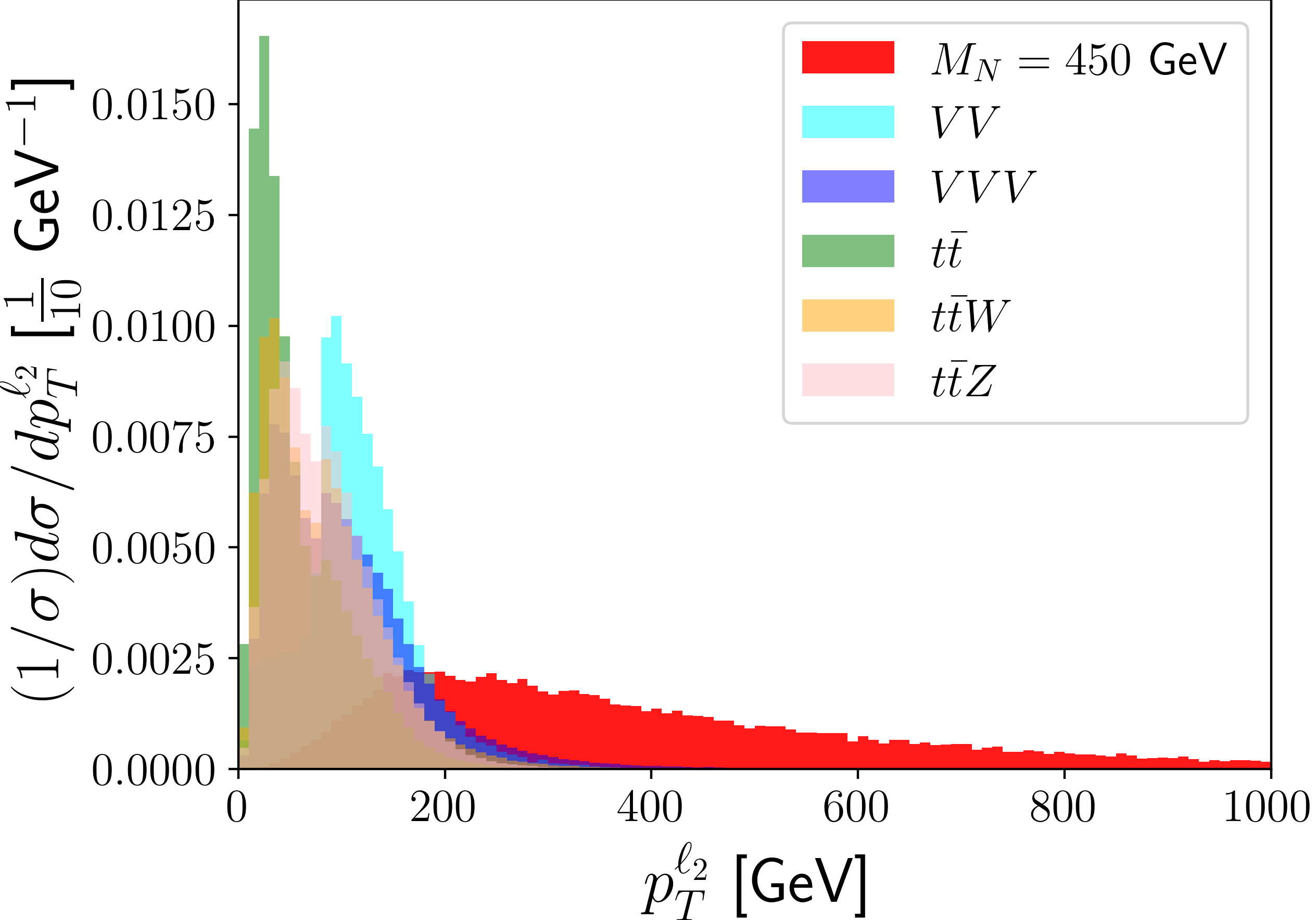}	
	\includegraphics[scale=0.37]{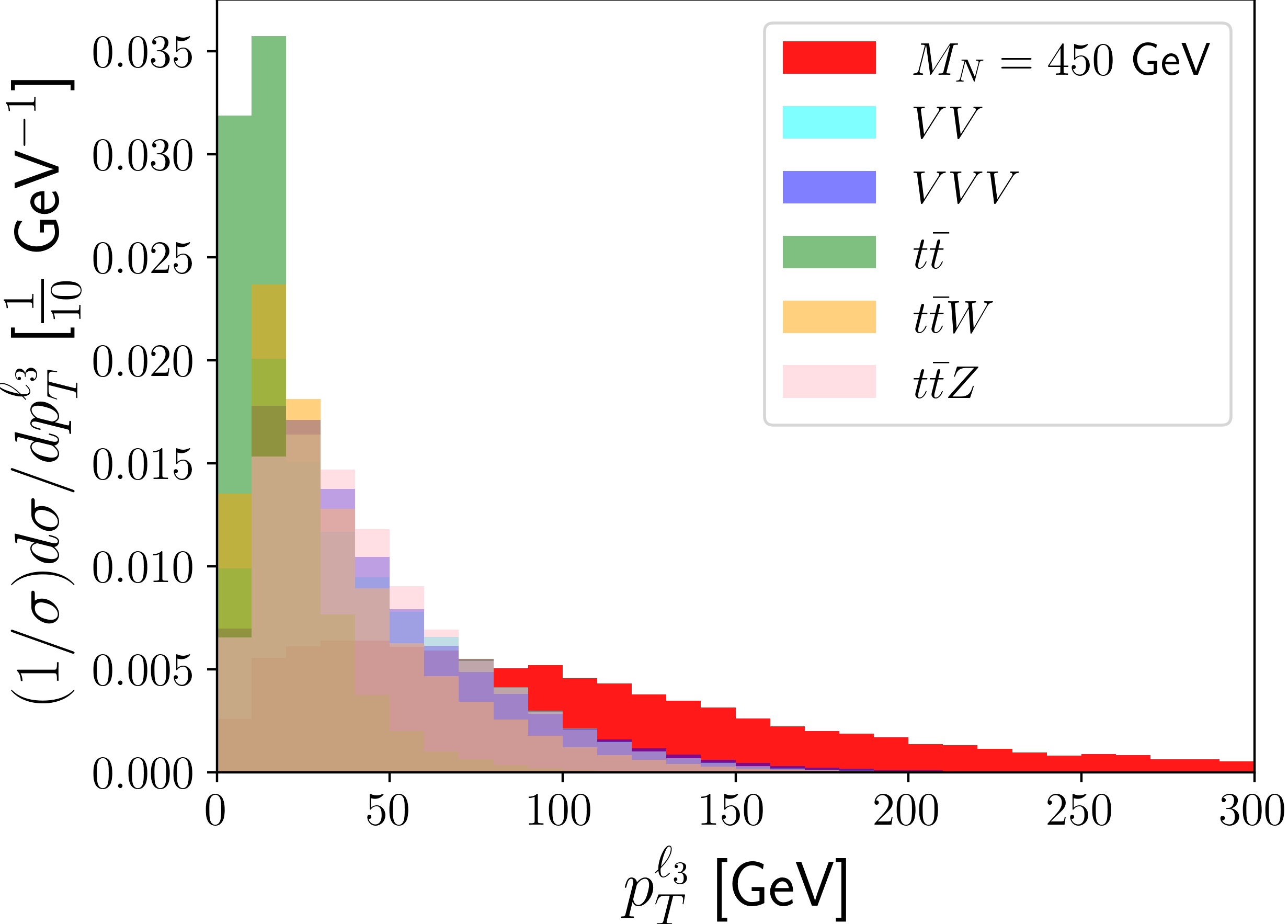}
	\includegraphics[scale=0.37]{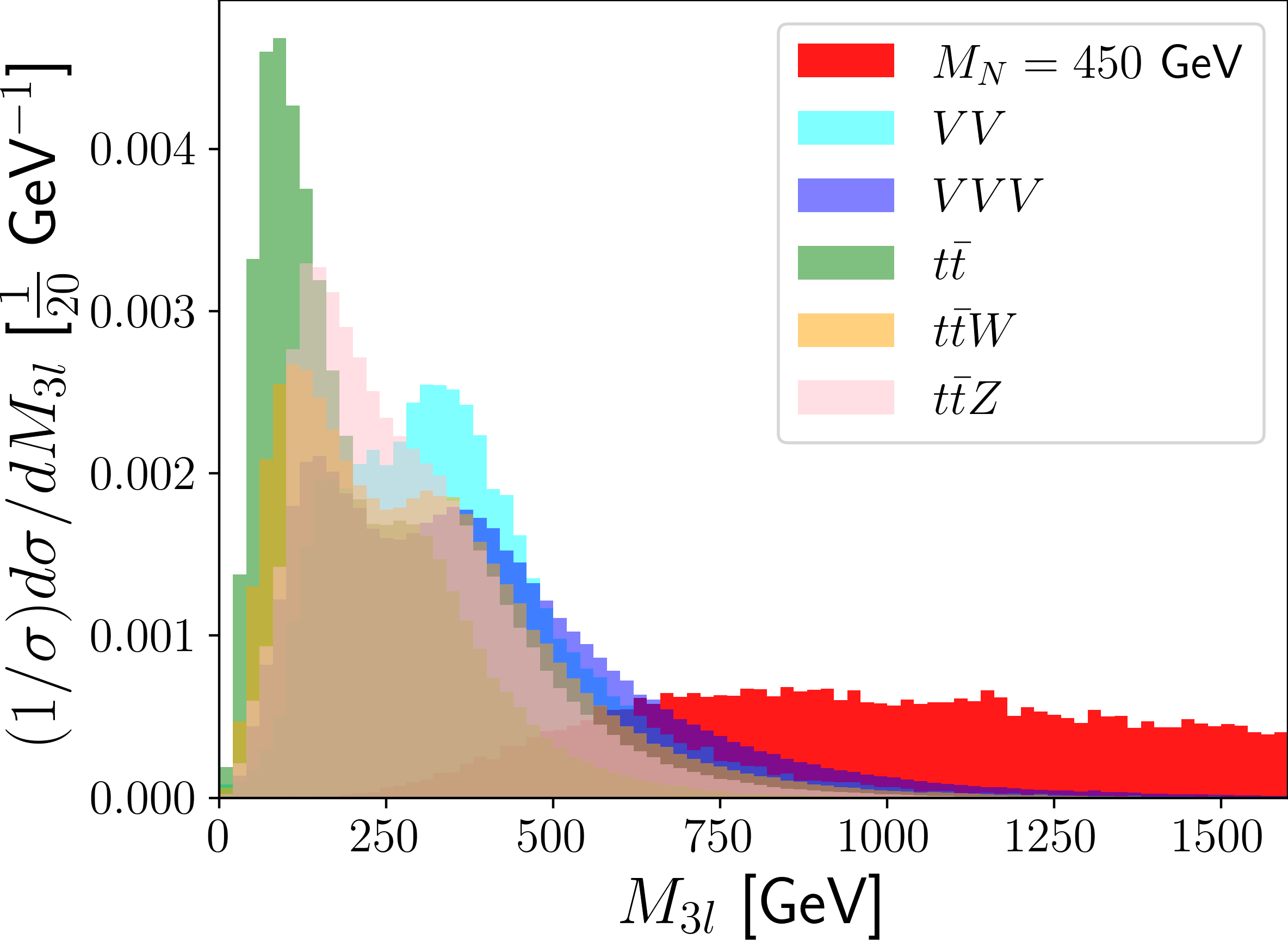}
	\includegraphics[scale=0.37]{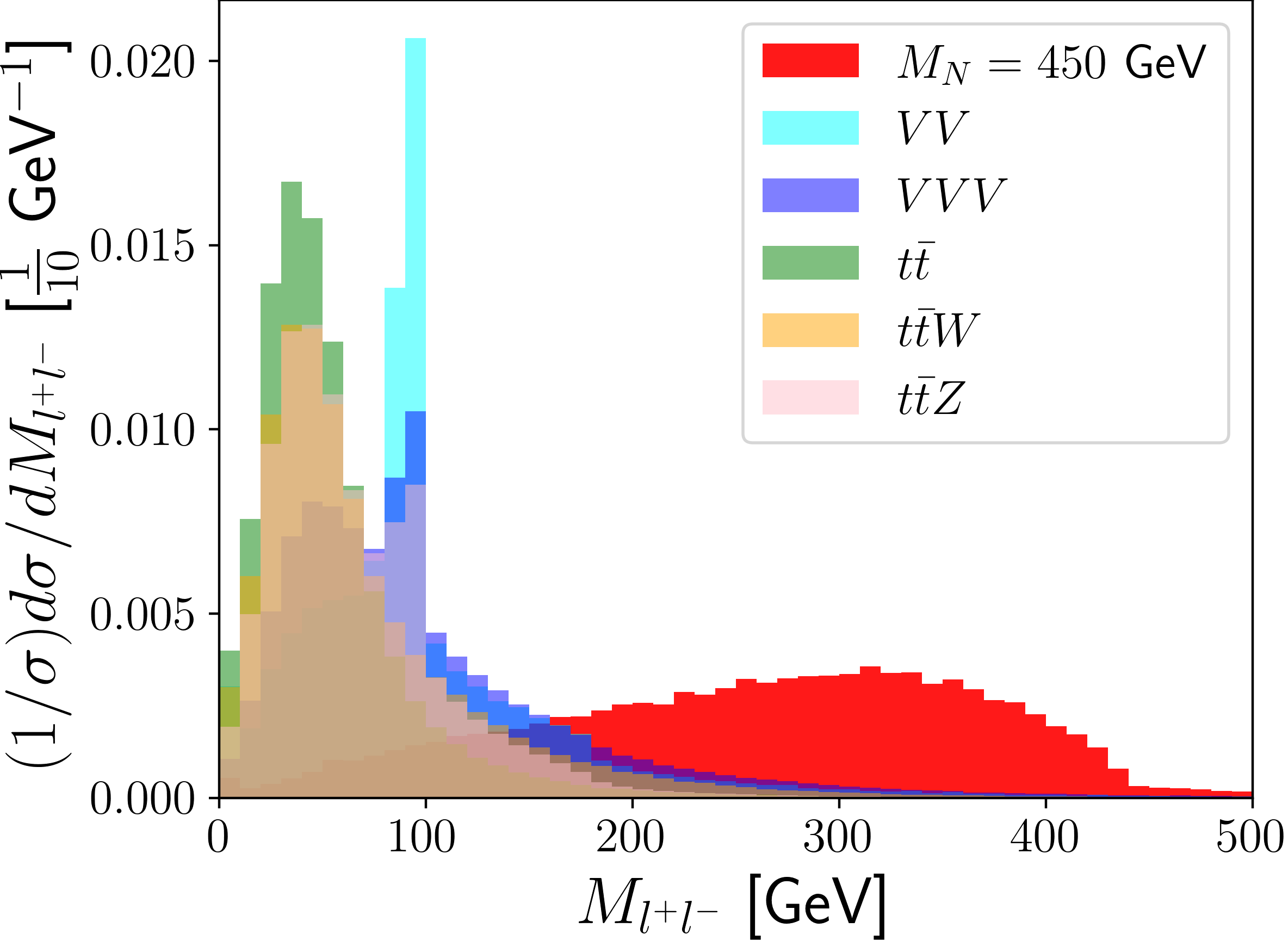}
	\includegraphics[scale=0.37]{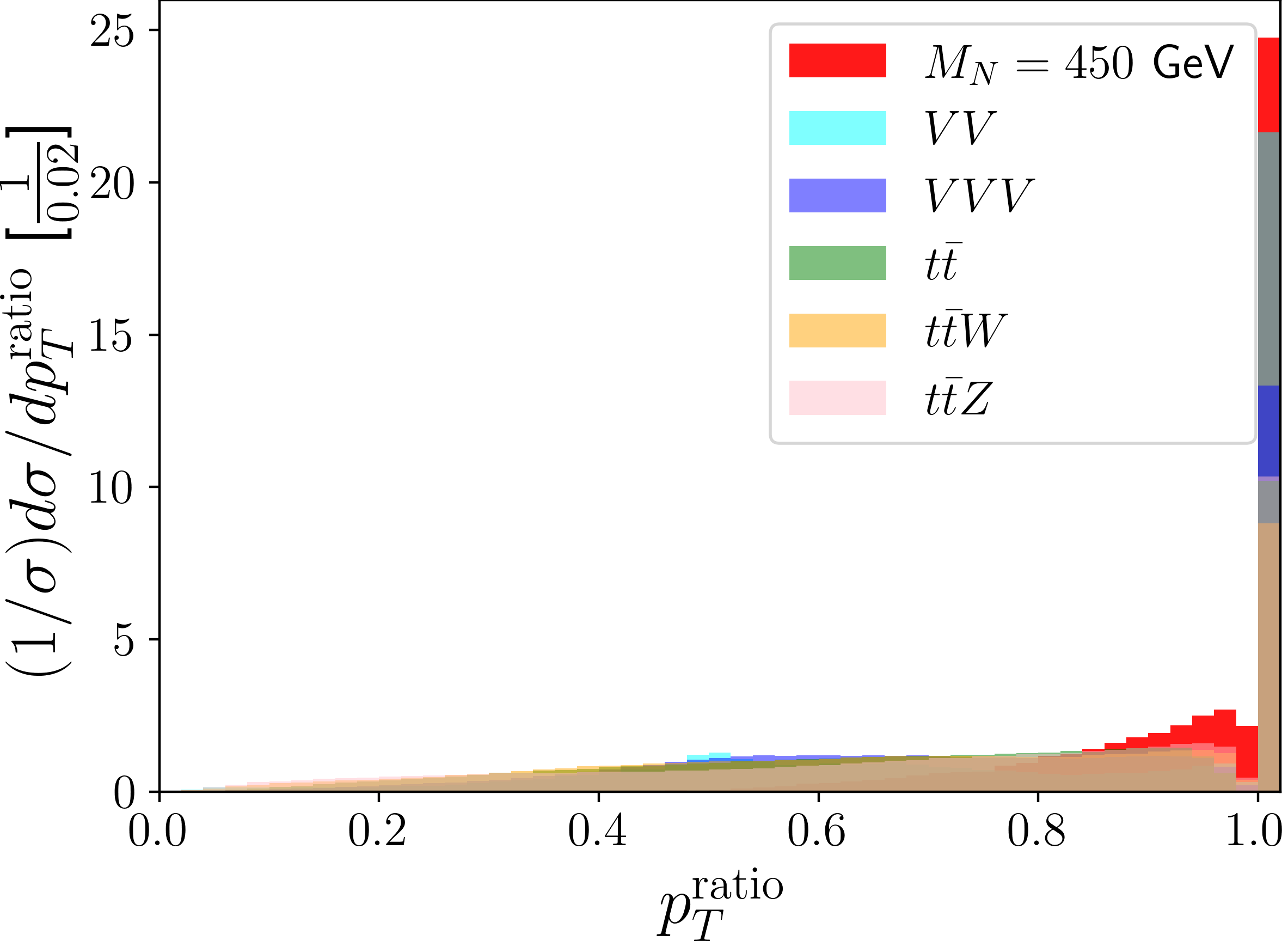}

\caption{The figure showcases different kinematic variables employed, with $M_N = 450$ GeV, $C_\Lambda = 4 \times 10^{-9}$, and $C^\prime_\Lambda = 4 \times 10^{-8}$.}
\label{Fig:dist_highmn_cpgtc}
\end{figure}

	In FIG.~\ref{Fig:dist_highmn_cpgtc}, we present the distributions for the variables used to analyse scenario d): $M_N=450 \text{ GeV } > M_{W}$ and $C_\Lambda < C^\prime_\Lambda $. Compared to the previous scenario b), $N$ is heavy and is moderately boosted, leading to well-isolated leptons. We follow the same isolation criteria as mentioned in section \ref{sec:5}.  The $p_T$ requirement of the leading three leptons are: $p_T^{\ell_1} > 350$ GeV, $p_T^{\ell_2} > 150$ GeV, $p_T^{\ell_3} > 50$ GeV. The variables with better discrimination power between the signal and SM background are $M_{3\ell},M_{\ell^+ \ell^-} \text{ and } p_T^{\mathrm{ratio}}$. $p_T^{\mathrm{ratio}}$ is defined as the ratio of sum of $p_T$ of all selected electrons and sum of $p_T$ of all selected leptons, both muons and electrons.  Our signal has at least two electrons out of which one electron is directly produced from four-Fermi interaction and hence has larger momentum. As a result, $p_T^{\mathrm{ratio}}$ for the signal populates towards larger values and helps to distinguish the signal from the SM background. After scrutinising the distributions as depicted in FIG.~\ref{Fig:dist_highmn_cpgtc}, we impose cuts on the discriminating variables to maximise the signal-to-background ratio, listed as:
$M_{\ell^\pm \ell^\mp}  > 100$ GeV, $M_{3 \ell} > 550$ GeV and $p_T^{\mathrm{ratio}} > 0.75$. 

Since the distributions for $C_\Lambda > C^\prime_\Lambda$ case, i.e. $C_\Lambda = 7 \times 10^{-8} \text{ and } C^\prime_\Lambda = 7 \times 10^{-9}$ is similar to FIG.~\ref{Fig:dist_highmn_cpgtc}, hence we do not show them explicitly.For the case \( C_{\Lambda} > C_{\Lambda}^{\prime} \), when we choose a heavier \( N \) with \( M_N = 450 \) \, \text{GeV}, which is greater than \( M_N^{\text{th}_2} = 86 \) \, \text{GeV}, the four-Fermi interaction dominates the production cross-section. This behavior is similar to the \( C_{\Lambda} < C_{\Lambda}^{\prime} \) case, resulting in comparable kinematic distributions as previously mentioned. Hence,  in this case, we implement the same set of cuts as performed for  $C_\Lambda < C^\prime_\Lambda$. The cut-flow for both the scenarios is enlisted in TABLE.~\ref{tab:cut_flow_mngtmw_redi}. The description of $no.~\ell^\pm \geq 3$ cut is similar to before. We can achieve a significance greater than $4\sigma$ for $C_\Lambda < C^\prime_\Lambda$. For $C_\Lambda < C^\prime_\Lambda$, we achieve less than $2\sigma$ significance. For two benchmark points, namely $M_N = 50$ GeV with $C_\Lambda < C^\prime_\Lambda$ and $M_N = 450$ GeV with $C_\Lambda > C^\prime_\Lambda$, we achieve less than $2\sigma$ significance using cut-based analysis. Consequently, we apply BDT analysis to these cases.

\begin{table*}[b!]
	\centering
	\small 
	\begin{adjustbox}{width=0.8\textwidth} 
		\addtolength{\tabcolsep}{-1pt}
		\begin{tabular}{||c|c|c|c|c|c|c|c||}
			\hline 
			& no. bjets = 0 & no. $\ell^\pm \geq 3$  & $M_{\ell^\pm \ell^\mp} > 100$ &  $M_{3\ell} > 550$ &$p_{T_{\mathrm{ratio}}} > 0.75$& ${\sigma}_{\text{eff}}$[fb] & ${\eta}_{s}$\\
			\hline 
			450 GeV ($C_{\Lambda} > C_{\Lambda}^{\prime}$)  [2.2 $\times 10^{-2}$ fb] & 2.2 $\times 10^{-2}$ & 6.58 $\times 10^{-3}$ & 6.52 $\times 10^{-3}$ & 6.52 $\times 10^{-3}$ & 5.98 $\times 10^{-3}$ & 5.98 $\times 10^{-3}$ & {1.45}\\
			450 GeV ($C_{\Lambda} < C_{\Lambda}^{\prime}$)  [7.6 $\times 10^{-2}$ fb] & 7.6 $\times 10^{-2}$ & 2.28 $\times 10^{-2}$ & 2.27 $\times 10^{-2}$ & 2.26 $\times 10^{-2}$ & 2.08 $\times 10^{-2}$ & 2.08 $\times 10^{-2}$ & {4.49}\\
			\hline \hline
			$VV$ [9847 fb]  & 9847 & 2.18 $\times 10^{-1}$ & 9.54 $\times 10^{-2}$ & 7.16 $\times 10^{-2}$ & 4.26 $\times 10^{-2}$ & 4.26 $\times 10^{-2}$ &\\
			$VVV$ [5.53 fb] & 5.53 & 7.55 $\times 10^{-3}$ & 3.98 $\times 10^{-3}$ & 3.85 $\times 10^{-3}$ & 2.01 $\times 10^{-3}$ & 2.01 $\times 10^{-3}$&\\
			$t\bar{t}$ [51170 fb] & 11279.98 & 0 & 0 & 0 & 0 & 0 &\\
			$t\bar{t}W^\pm$  [13.22 fb]  & 2.30 & 5.55 $\times 10^{-4}$  & 4.76 $\times 10^{-4}$  & 4.76 $\times 10^{-4}$  & 1.85 $\times 10^{-4}$  & 1.85 $\times 10^{-4}$&\\
			$t\bar{t}Z$  [7.26 fb] & 1.31 & 1.23 $\times 10^{-3}$  & 5.52 $\times 10^{-4}$  & 3.92 $\times 10^{-4}$  & 1.89 $\times 10^{-4}$  & 1.89 $\times 10^{-4}$ &\\
			\hline 
		\end{tabular}
	\end{adjustbox}
		\caption{Cross-sections of signal and backgrounds after imposing our proposed sets of cuts, considering $M_N = 450 $ GeV and $C_{\Lambda} =4 \times 10^{-9}(7 \times 10^{-8})  $, $C_{\Lambda}^{\prime} = 4 \times 10^{-8}( 7 \times 10^{-9})$.}
		\label{tab:cut_flow_mngtmw_redi}
\end{table*}

In addition to the kinematic variables utilised in the cut-based analysis, we incorporate several additional variables with robust background-signal discriminating power for the multivariate analysis. These variables are 
${p_T}^{\mathrm{ratio}}_{\mathrm{miss},3\ell}$, $\sum_{i=1}^{2} \Delta \phi(\mathrm{miss}, \ell_{i})$, ${\prod}_{i=1}^{2}{\Delta \phi}(\mathrm{miss},{\ell}_{i})$,   $M_{T}^{{\ell}_2}$, ${\Delta \phi}^{\mathrm{miss}}_{3{\ell}}$ and ${\Delta \phi }^{\mathrm{miss}}_{\ell}$.   ${p_T}^{\mathrm{ratio}}_{\mathrm{miss},3\ell}$ represents the ratio of missing transverse momentum to the sum of the transverse momentum of the first three leading leptons, $\sum_{i=1}^{2} \Delta \phi(\mathrm{miss}, \ell_{i})$ (${\prod}_{i=1}^{2}{\Delta \phi}(\mathrm{miss},{\ell}_{i})$) is the sum (product) of the differences in azimuthal angle ($\Delta \phi$) between the missing energy and the first two leading leptons. $M_{T}^{{\ell}_2}$ corresponds to the transverse mass of the second leading lepton and the missing energy, sorted against the mass of the W-boson. The other variables, such as, ${\Delta \phi}^{\mathrm{miss}}_{3{\ell}}$ is the difference in azimuthal angle for the system composed of the first three leading leptons and the missing energy, and  ${\Delta \phi }^{\mathrm{miss}}_{\ell}$ implies the maximum difference in azimuthal angle between the missing energy and the leptons not included in the $M_{T}^{{{\ell}^+}{{\ell}^-}}$ calculation.

For \(M_N = 50\) GeV with \(C_\Lambda < C^\prime_\Lambda\), the significance reaches \(4.14\sigma\), while for \(M_N = 450\) GeV with \(C_\Lambda > C^\prime_\Lambda\), it is \(3.16\sigma\), based on the BDT analysis.

\subsection{Results and discussion for varying $M_N$}

This analysis underscores the intricate balance between different cuts and their impact on signal cut efficiency, particularly in scenarios where the mass of the RHN relative to the $W$ boson mass is a determining factor.
\begin{figure}[htb!]
	\centering
	\includegraphics[scale=0.55]{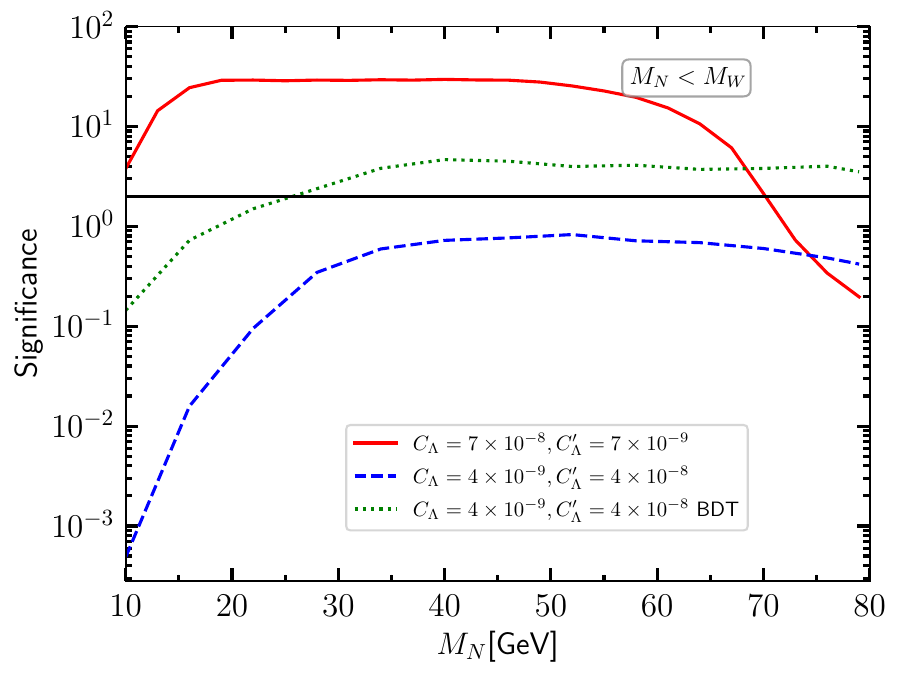}
	\includegraphics[scale=0.55]{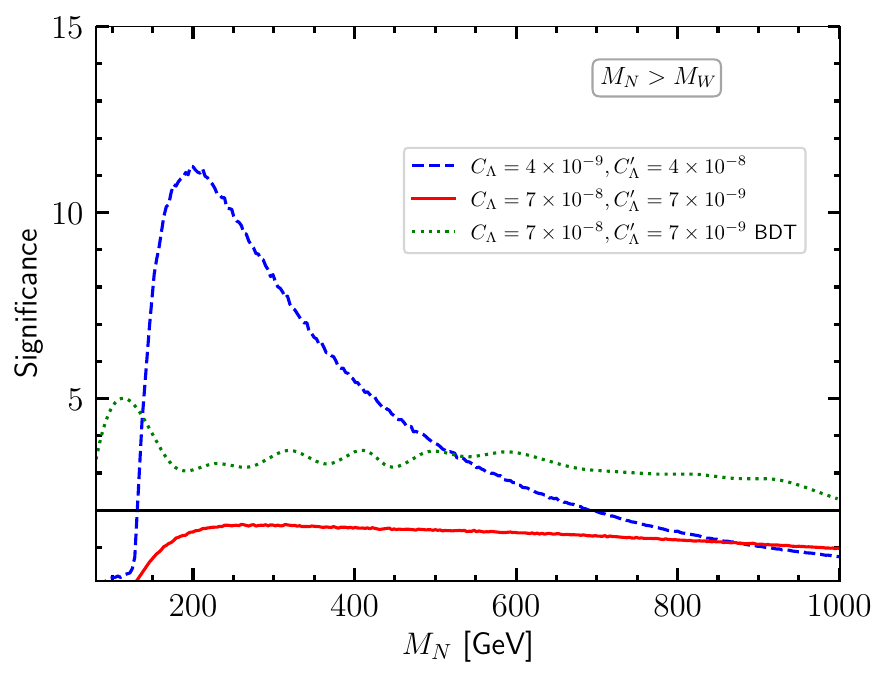}

	\caption{Variation of signal significance $\eta_s$ with $M_N$ at HL-LHC for $\sqrt{s} =14$ TeV. c.m.energy and $\mathcal{L} = 3000 \text{ fb}^{-1}$. In the left and right panel, we show the results for $M_N < M_W $ and $M_N > M_W $, respectively. The red and blue  lines represent the significance obtained for $C_{\Lambda}> C_{\Lambda}^{\prime}$  and  $C_{\Lambda}< C_{\Lambda}^{\prime}$ scenario using cut-based analysis. The green line represents achieved significance  for  $C_{\Lambda}< C_{\Lambda}^{\prime}$ scenario using BDT analysis. The black line represents $2 \sigma$ significance.}
	\label{Fig:significance_HL_LHC}
\end{figure}
The signal significance ${\eta}_{s}$ is computed using relation: 
\begin{equation}
{\eta}_{s} = \frac{S}{\sqrt{S+B}} \times \sqrt{\mathcal{L}}, 
\label{eq:significance}
\end{equation} where $S = \sigma_s \times \epsilon_s$ represents the product of the signal cross section $\sigma_s$ and its corresponding cut efficiency $\epsilon_s$, and $B = \sum_{i}\sigma_{b}^{i} \times \epsilon_{b}^{i}$ denotes the summation over background cross sections $\sigma_{b}^{i}$ multiplied by their respective cut efficiencies $\epsilon_{b}^{i}$, for each of the background $i$. We represent $\sigma_{s}$ and $\sigma_{b}^{i}$ in femtobarns (fb) and luminosity $\mathcal{L}$ in $\text{fb}^{-1}$.

In the left panel of FIG.~\ref{Fig:significance_HL_LHC}, the significance for benchmark values of $C_{\Lambda}$ and $C_{\Lambda}^{\prime}$ are shown for $M_{N} < M_{W}$. To obtain the significance, we followed the cut-based analysis mentioned in the TABLE~\ref{tab:CMSlowmass} and TABLE~\ref{tab:cut_flow_mnlwmw_redi}, respectively, along with the BDT analysis for $C_{\Lambda} < C_{\Lambda}^{\prime}$.
	
For $C_{\Lambda} > C_{\Lambda}^{\prime}$ case, for most of the mass points, we get more than $5 \sigma$ significance. For $C_{\Lambda} < C_{\Lambda}^{\prime} $, the cut-based method fails to yield more than a $2\sigma$ significance. Thus, we employ multivariate analysis (BDT), shown by the green line, which reflects a significant improvement in significance, surpassing $2 \sigma$ for most of the mass points.

In the right panel of FIG.~\ref{Fig:significance_HL_LHC}, the significance for benchmark values of $C_{\Lambda}$ and $C_{\Lambda}^{\prime}$ are shown for $M_{N} > M_{W}$. To evaluate the significance, we followed the cut-based analysis mentioned in TABLE~\ref{tab:cut_flow_mngtmw_redi}, along with the BDT analysis for $C_{\Lambda} > C_{\Lambda}^{\prime}$. 

For $C_{\Lambda} < C_{\Lambda}^{\prime} $, we achieve more than $3 \sigma$ significance for most of the considered mass points. However, when considering $C_{\Lambda} > C_{\Lambda}^{\prime} $, the significance falls below $2\sigma$ for all the mass points under consideration. Thus, we employ multivariate analysis (BDT) for this case, shown by the green line, which reflects a significant improvement in significance, surpassing $3 \sigma$ for almost all the mass points.

\begin{figure}[htb!]
	\centering
	
	\includegraphics[scale=0.55]{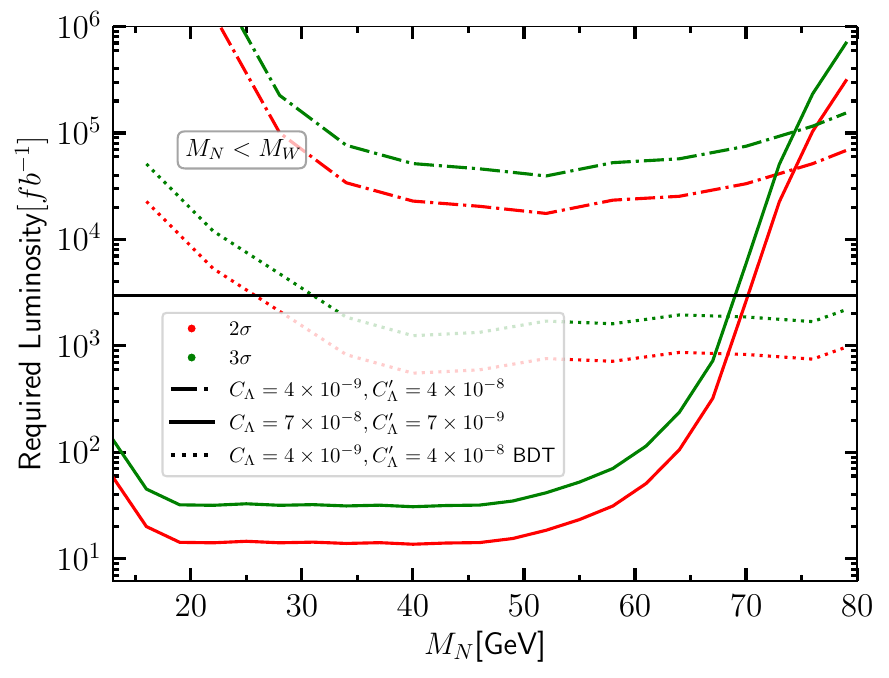}
	\includegraphics[scale=0.55]{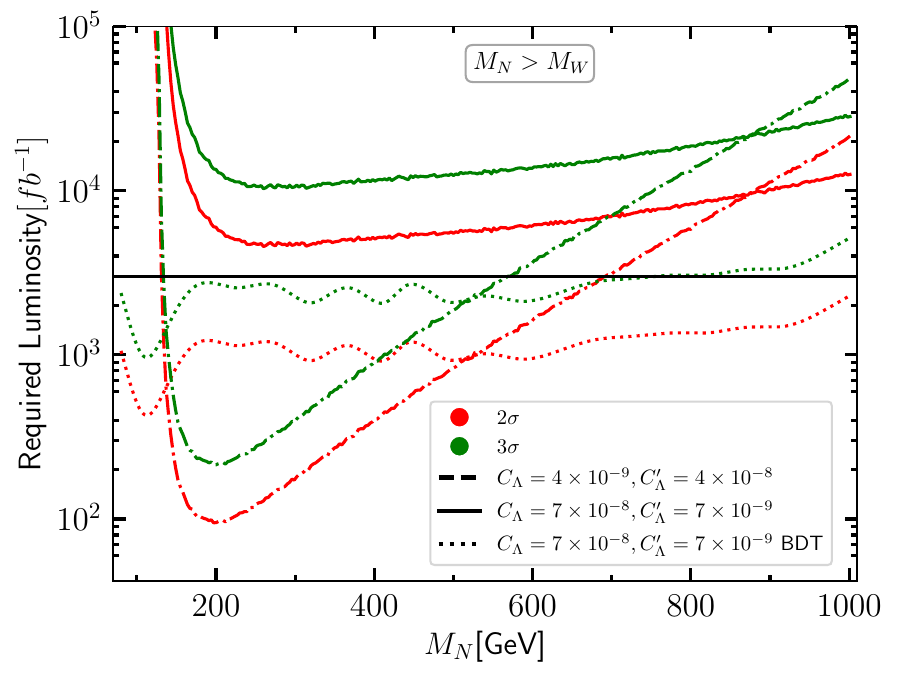}
	\caption{Required luminosity  to achieve  $2 \sigma$(red), $3 \sigma$(green) and $5 \sigma$(blue) statistical significance, for different values of RHN  mass. In left (right) we show the results for $M_N < M_W(M_N > M_W)$ case. The black line represents $3000\,  \text{fb}^{-1}$ luminosity. The dotted lines represent  result obtained with BDT analysis. The solid and dot-dashed lines represent the results obtained with cut-based analysis.}
	\label{Fig:luminosity_reqired}
\end{figure}

In the left panel of FIG.~\ref{Fig:luminosity_reqired}, we depict the required luminosity to attain $2 \sigma$ and $3 \sigma$ significance levels for our benchmark values of $C_{\Lambda}$ and $C_{\Lambda}^{\prime}$, for $M_{N} < M_{W}$. With the exception of the cut-based analysis for \(C_{\Lambda} < C_{\Lambda}' \), less than \(3000 \, \text{fb}^{-1}\) luminosity is sufficient to achieve both \(2\sigma\) and \(3\sigma\) significance for most of the mass points. In the right panel of FIG.~\ref{Fig:luminosity_reqired}, we illustrate the necessary luminosity required to achieve $2 \sigma$ and $3 \sigma$ significance levels for our benchmark values of $C_{\Lambda}$ and $C_{\Lambda}^{\prime}$, focusing on the scenario where $M_{N} > M_{W}$. Except for the cut-based analysis in the $C_{\Lambda} > C_{\Lambda}'$ scenario, reaching a significance level of $2\sigma$ for most of the mass points requires less than $3000$ fb$^{-1}$ of luminosity.

\section{Conclusion \label{conclusion}}

We investigate the impact of  $d=6$ $N_R$-EFT operators $\mathcal{O}_{HNe}$ and $\mathcal{O}_{duNe}$ on tri-lepton and missing energy signals at the HL-LHC. We primarily focus on parameter spaces yielding at least three leptons, among which at least two are electrons/positrons. The operators mentioned above enhance non-trivial decay modes. The large Wilson coefficient of $\mathcal{O}_{duNe}$ operator substantially contributes to $p p \to eN$ production. The operator $\mathcal{O}_{HNe}$ modifies the $W-e-N$ vertex and can impact the $p p \to e N$ cross-section via $W$ mediation. We consider four different scenarios, (a) \(M_N < M_W\), \(C_{\Lambda} > C^{\prime}_{\Lambda}\); (b) \(M_N < M_W\), \(C_{\Lambda} < C^{\prime}_{\Lambda}\); (c) \(M_N > M_W\), \(C_{\Lambda} > C^{\prime}_{\Lambda}\); and (d) \(M_N > M_W\), \(C_{\Lambda} < C^{\prime}_{\Lambda}\).

Inspired by \cite{CMS:2018iaf}, we employ a CMS-motivated cut-based analysis for all the above scenarios. However, this approach yields significant results only for scenario (a). For a specific mass point \(M_N = 50 \text{ GeV} < M_W\) in scenario (b), we find that the isolation requirement together with the \(M_{3\ell}\) cut, as demanded in \cite{CMS:2018iaf}, removes most of the signal samples. Therefore, we modify the lepton isolation criterion. For scenarios (c) and (d) with \(M_N = 450 \text{ GeV} > M_W\), we find that the kinematic variables such as \(M_{\ell^+ \ell^-}\) and \(p^{\mathrm{miss}}_T\) are insufficient to significantly reduce the SM background. Consequently, we introduce new variables: \(\Delta \phi_{\mathrm{ratio}}\) and \(p^{\mathrm{ratio}}_T\). Along with these, we also use variables such as \(M_{3\ell}\) and \(M^{\ell^+ \ell^-}_T\). In scenario (d), this approach enhances the signal significance relative to the CMS-motivated analysis, leading to more than \(5\sigma\) significance for a right-handed neutrino mass in the range of 200-400 GeV. For the remaining scenarios (b) and (c), the significance remains somewhat low. Therefore, we conduct a multivariate analysis using 20 variables, including some newly proposed ones, to improve the significance. This further enhances the signal significance, achieving more than \(2\sigma\) significance for scenario (c). For scenario (b), we achieve more than \(3\sigma\) significance for \(M_N > 35\) GeV.

In conclusion, incorporating ratio variables has proven highly effective, capturing correlations between different parameters while aligning with the signal topology. This approach offers promising prospects, even in small signal cross-section scenarios. We can significantly improve signal sensitivity by leveraging Boosted Decision Trees (BDT) and employing carefully selected discriminating variables. As part of future endeavours, exploring alternative machine-learning techniques for classification purposes could yield valuable insights and enhance the robustness of our approach.

\acknowledgments 

The authors acknowledge the use of SAMKHYA: High-Performance Computing Facility provided by the Institute of Physics (IOP), Bhubaneswar. M.M acknowledges the IPPP Diva Award research grant. M.T. is supported by the Fundamental Research Funds for the Central Universities, the One Hundred Talent Program of Sun Yat-sen University, China, and by the JSPS KAKENHI Grant, the Grant-in-Aid for Scientific Research C, Grant No.~18K03611. The authors thank A. Sarkar and S.P. Maharathy for their initial contributions to the project. 

\section*{A.  Appendix  \label{Appendix_n}}

We provide the analytic expressions of two body decays $N \to e W$, $N \to \nu_e Z, \nu_e H$ along with the relevant three body decays
$N \to e u \bar{d}, \nu_e d \bar{d}, \nu_e u \bar{u}, \nu_e \ell^+ \ell^-, \nu_e \nu \bar{\nu}$. 

\subsection*{ A.1.  2 body decay widths}
The two body decay modes of $N \rightarrow ab$ are as follows,

\begin{align}
	\Gamma (N \rightarrow e W) &= \frac{g^{2}}{64\pi M_{N}M_{W}^{2}} \bigg[ \left(\chi^{2} + A^{2} \right)  \left( M_{W}^{2}(M_{{e}}^{2} + M_{N}^{2}) + (M_{e}^{2} - M_{N}^{2})^{2} - 2 M_{W}^{4} \right) \nonumber \\  &- 12 \left(\chi. A \right)M_{e} M_{N} M_{W}^{2} \bigg]  \lambda^{\frac{1}{2}}(1,x_{e}^{2},x_{W}^{2}),
\end{align}
Where,
 $ A = \left( \frac{ C_{\Lambda} v^2}{2 } \right) $ and $ \lambda(a,b,c) = a^2 + b^2 + c^2 - 2ab - 2bc - 2ac. $ , $x_{i} = \frac{M_{i}^{2}}{M_{N}^{2}}$

In this paper, we have focused exclusively on the $N \rightarrow e W$ decay. As demonstrated, this decay mode is influenced directly by the active-sterile mixing parameter $\chi$ and the Wilson coefficient $C_{\Lambda}$, with interference occurring among them.

\begin{align}
	\Gamma (N \rightarrow \nu_{e} H) &= \frac{\chi^{2}}{32 \pi M_{N} v^{2}} (M_{N}^{2} - M_{H}^{2})^{2},
\end{align}
and
\begin{align}
	\Gamma(N \rightarrow {\nu}_{e} Z) = \frac{(M_{N}^{2} - M_{Z}^{2})^2}{128 \pi c_{w}^{2}M_{Z}^{2}M_N^3} (g^2 \chi^{2}(M_N^2 + 2 M_{Z}^{2}))
\end{align}
The $N \rightarrow \nu_{e} H$ and $N \rightarrow \nu_{e} Z$ decay modes receive contributions solely from $\chi$.

\subsection*{A.2.  3 body decay widths	\label{threebody}}
The three body decay modes of $N \rightarrow abc$ are as follows,

\begin{align}
	\Gamma(N \rightarrow e u_{\alpha} \bar{d_{\beta}}) &= \frac{M_N^5 N_{C}}{512 \pi^3} \left( 
	4 \left[ \chi^2 \left( \frac{g^2 V_{\text{CKM}}^{\alpha \beta}}{2 M_W^2} \right)^2 + \left(C_{\Lambda}^{\prime} \right)^2 \right] I_1(x_{e}, x_{u_{\alpha}}, x_{d_{\beta}}) \right.  + 8 \chi A   \left( \frac{g^2 V_{\text{CKM}}^{\alpha \beta}}{2 M_W^2} \right)^2  I_2(x_{u_{\alpha}}, x_{d_{\beta}}, x_{e}) \nonumber \\
	&\quad + 4 \left[ A^2 \left( \frac{g^2 V_{\text{CKM}}^{\alpha \beta}}{2 M_W^2} \right)^2 \right] I_1(x_{e}, x_{d_{\beta}}, x_{u_{\alpha}})  - 8 \left[ A \left( \frac{g^2 V_{\text{CKM}}^{\alpha \beta}}{2 M_W^2} \right) C_{\Lambda}^{\prime} \right] G_3(x_{u_{\alpha}}, x_{d_{\beta}}, x_{e}) \nonumber \\
	&\quad \left. + 32 \chi \left[ \left( \frac{g^2 V_{\text{CKM}}^{\alpha \beta}}{2 M_W^2} \right) C_{\Lambda}^{\prime} \right] I_5(x_{u_{\alpha}}, x_{d_{\beta}}, x_{e})
	\right)
\end{align}
Where, 
$$I_1(x_a, x_b, x_c) = \int_{(x_a + x_b)^2}^{(1 - x_c)^2} \frac{(z - x_a^2 - x_b^2) \cdot (1 + x_c^2 - z) \cdot \sqrt{\lambda(1, z, x_c^2)} \cdot \sqrt{\lambda(z, x_a^2, x_b^2)}}{z} \, dz
$$
$$I_2(x_a, x_b, x_c) = -\int_{(x_a + x_b)^2}^{(1 - x_c)^2} \frac{(z - x_a^2 - x_b^2) \cdot x_c \cdot \sqrt{\lambda(1, z, x_c^2)} \cdot \sqrt{\lambda(1, x_a^2, x_b^2)}}{z} \, dz
$$

$$I_5(x_a, x_b, x_c) = -\int_{(x_a + x_b)^2}^{(1 - x_c)^2} \frac{x_a \cdot x_b \cdot x_c \cdot \sqrt{\lambda(1, z, x_c^2)} \cdot \sqrt{\lambda(1, x_a^2, x_b^2)}}{z} \, dz
$$

$$G_3(x_a, x_b, x_c) = -\int_{(x_a + x_b)^2}^{(1 - x_c)^2} \frac{x_a \cdot x_b \cdot (1 + x_c^2 - z) \cdot \sqrt{\lambda(1, z, x_c^2)} \cdot \sqrt{\lambda(z, x_a^2, x_b^2)}}{z} \, dz
$$
$u_{\alpha}$ and $d_{\beta}$ correspond to up- and down-type light quarks, respectively.

\begin{align}
	\Gamma(N \rightarrow \nu_e d_{\beta} \bar{d_{\beta}}) &= 
	\frac{M_N^5}{512 \pi^3}  \Bigg(
	\frac{g^4  \chi^2  (g_L^2 + g_R^2)}{4 M_W^4}  I_1(x_{\nu_e}, x_{d_{\beta}}, x_{d_{\beta}})  - \frac{g_L g_R  g^4  \chi^2}{2 M_W^4}  G_3(x_{d_{\beta}}, x_{d_{\beta}}, x_{\nu_e}) \nonumber \\
	&\quad   + \frac{Y_{\nu}^2  Y_{d_{\beta}}^2}{2 M_H^4} \left[ I_1(x_{d_{\beta}}, x_{d_{\beta}}, x_{\nu_e}) + 2 G_3(x_{d_{\beta}}, x_{d_{\beta}}, x_{\nu_e}) \right] 
	\Bigg),
\end{align}
 Where $g_L = \frac{2}{3} s_{w}^{2} -1$, $g_R = \frac{2}{3} s_{w}^{2}$, $Y_{\nu} = \frac{\sqrt{2} \chi M_N}{v}$ and $Y_{d_{\beta}} = \frac{\sqrt{2} M_{d_{\beta}}}{v}$. $d_{\beta}$ stands for down-type light quarks.

\begin{align}
	\Gamma(N \rightarrow \nu_e u_{\alpha} \bar{u_{\alpha}}) &= 
	\frac{M_N^5}{512 \pi^3}  \Bigg(
	\frac{g^4  \chi^2  (g_L^2 + g_R^2)}{4 M_W^4}  I_1(x_{\nu_e}, x_{u_{\alpha}}, x_{u_{\alpha}})  - \frac{g_L g_R  g^4  \chi^2}{2 M_W^4}  G_3(x_{u_{\alpha}}, x_{u_{\alpha}}, x_{\nu_e}) \nonumber \\
	&\quad  + \frac{Y_{\nu}^2  Y_{u_{\alpha}}^2}{ 2 M_H^4} \left[ I_1(x_{u_{\alpha}}, x_{u_{\alpha}}, x_{\nu_e}) + 2 G_3(x_{u_{\alpha}}, x_{u_{\alpha}}, x_{\nu_e}) \right]
	\Bigg),
\end{align}
Where $g_L = 1- \frac{4}{3} s_w^2$, $g_R = - \frac{4}{3} s_w^2$ and $Y_{u_{\alpha}} = \frac{\sqrt{2} M_{u_{\alpha}}}{v}$. $u_{\alpha}$ represents for up-type light quarks.

\begin{align}
	\Gamma(N \rightarrow \nu_e \ell_{k}^{+} \ell_{k}^{-}) &= 
	\frac{M_N^5}{512 \pi^3}  \Bigg(
	\frac{g^4  \chi^2  (g_L^2 + g_R^2)}{4 M_W^4}  I_1(x_{\nu_e}, x_{\ell_{k}}, x_{\ell_{k}})  - \frac{g_L g_R  g^4  \chi^2}{2 M_W^4}  G_3(x_{\ell_{k}}, x_{\ell_{k}}, x_{\nu_e}) \nonumber \\
	&\quad   + \frac{Y_{\nu}^2  Y_{\ell_{k}}^2}{ 2 M_H^4} \left[ I_1(x_{\ell_{k}}, x_{\ell_{k}}, x_{\nu_e}) + 2 G_3(x_{\ell_{k}}, x_{\ell_{k}}, x_{\nu_e}) \right] 
	\Bigg),
\end{align}
Where $g_L = 2s_w^2 - 1$, $g_R = 2s_w^2$ and $Y_{\ell_{k}} = \frac{\sqrt{2} M_{\ell_{k}}}{v}$.  $\ell_{k}$ represents the muon ($\mu$), and tau ($\tau$).

\begin{align}
	\Gamma(N \rightarrow \nu_e \nu_k \bar{\nu_k}) &= 
	\frac{M_N^5}{512 \pi^3}  \left(
	\frac{g^4  \chi^2  (g_L^2 + g_R^2)}{4 M_W^4}  I_1(x_{\nu_e}, x_{\nu_k}, x_{\nu_k})
	- \frac{g_L g_R  g^4  \chi^2}{2 M_W^4}  G_3(x_{\nu_k}, x_{\nu_k}, x_{\nu_e})
	\right),
\end{align}
Where $g_L = 1$ and $g_R = 0$ . $\nu_{k}$ refers to the  $\nu_{e}$, $\nu_{\mu}$ and $\nu_{\tau}$.
\begin{align}
	\Gamma(N \rightarrow e^{-} {\nu}_{k} \ell_{k}^{+}) &= 
	\frac{M_N^5 g^4}{512 \pi^3 M_W^4}  \left(
	A^2  I_1(x_{e}, x_{\ell_{k}}, x_{{\nu}_{k}}) \right.  + 2  A  \chi I_2(x_{{\nu}_{k}}, x_{\ell_{k}}, x_{e})  \left. +  \chi^2  I_1(x_{e}, x_{{\nu}_{k}}, x_{\ell_{k}})
	\right),
\end{align}

and
\begin{align}
	\Gamma(N \rightarrow \nu_e {e}^{-} {e}^{+}) &= 
	\frac{M_N^5}{512 \pi^3} \Bigg(  \frac{g^4 \chi^2 (g_L^2 + g_R^2)}{4 M_W^4} I_1(x_{\nu_e}, x_{e} x_{e}) - \frac{g_L g_R g^4 \chi^2}{2 M_W^4} G_3(x_{e}, x_{e}, x_{\nu_e}) \nonumber \\
	& + \frac{Y_\nu^2 Y_{e}^2}{2 M_H^4} \left(I_1(x_{e}, x_{e}, x_{\nu_e}) + 2 G_3(x_{e}, x_{e}, x_{\nu_e})\right) + \frac{g^4 A^2}{M_W^4} I_1(x_{e}, x_{e}, x_{\nu_e}) \nonumber \\
	&+ \frac{2 g^4 A \chi}{M_W^4} I_2(x_{\nu_e}, x_{e}, x_{e}) + \frac{g^4 \chi^2}{M_W^4} I_1(x_{e}, x_{\nu_e}, x_{e}) + 8 A C_{t2} C_{t3} (g_R - g_L) I_2(x_{\nu_e}, x_{e}, x_{e}) \nonumber \\
	&+ 8 A C_{t2} \frac{Y_\nu Y_{e}}{M_H^2} G_3(x_{e}, x_{e}, x_{\nu_e}) - 8 \chi g_L C_{t2} C_{t3} I_1(x_{e}, x_{\nu_e}, x_{e}) \nonumber \\
	& + 8 \chi g_R C_{t2} C_{t3} G_3(x_{e}, x_{e}, x_{\nu_e}) + 4 A C_{t_{2}} \frac{Y_{\nu} Y_{e}}{2 M_{H}^{2}}I_{1}(x_{e},x_{e}, x_{\nu_e}) \Bigg),
\end{align}

Where $
g_L = 2s_w^2 - 1$, $g_R = 2s_w^2$, $ 
Y_{e} = \frac{\sqrt{2} M_{e}}{v}$, $
C_{t2} = \frac{g^2}{2M_W^2}$ and $
C_{t3} = \frac{g^2 \chi}{4M_W^2}$. \\
As shown in the above expressions, the $N \rightarrow e u_{\alpha} \bar{d}_{\beta}$ decay mode is influenced by contributions from $\chi$, $C_{\Lambda}$, and $C_{\Lambda}^{\prime}$, with interference occurring among them. The $N \rightarrow \nu_e \nu_k \ell_{k}^{+}$ decay mode is affected by contributions from both $\chi$ and $C_{\Lambda}$. All other three-body decay modes are influenced solely by $\chi$.

\bibliographystyle{utphys}
\bibliography{bibliography} 
\end{document}